\documentclass[10pt,journal,compsoc]{IEEEtran}
\newif\ifpeerreview

\peerreviewfalse

\usepackage[nocompress]{cite}
\usepackage{url}
\usepackage{amsmath,amssymb,graphicx}

\usepackage{lipsum} 

\usepackage[switch]{lineno}

\usepackage{tabularx}
\usepackage{booktabs}
\usepackage{multirow}
\usepackage{hhline}
\usepackage{colortbl}
\usepackage{xcolor}
\usepackage[hidelinks]{hyperref}

\let\titleold\title
\renewcommand{\title}[1]{\titleold{#1}\newcommand{\thetitle}{#1}}
\def\maketitlesupplementary
   {
   \newpage
       \twocolumn[
        \centering
        \Large
        \textbf{\thetitle}\\
        \vspace{0.5em}Supplementary Material \\
        \vspace{1.0em}
       ] 
   }

\newcommand{\paperID}{20}

\title{Examining Joint Demosaicing and Denoising for
Single-, Quad-, and Nona-Bayer Patterns}

\author{SaiKiran Tedla$^\texttt{1,2}$, Abhijith Punnappurath$^\texttt{1}$, Luxi Zhao$^\texttt{1}$, Michael S. Brown$^\texttt{1,2}$\\
$^\texttt{1}$ AI Center-Toronto, Samsung Electronics\hspace{0.4cm} $^\texttt{2}$ York University 
{\tt\small {\{s.tedla,abhijith.p,lucy.zhao,michael.b1\}@samsung.com}\hspace{0.4cm}\tt\small {\{tedlasai,mbrown\}@yorku.ca}}
}

\begin{document}

\IEEEtitleabstractindextext{%

\begin{abstract}

  Camera sensors have color filters arranged in a mosaic layout, traditionally following the Bayer pattern. Demosaicing is a critical step camera hardware applies to obtain a full-channel RGB image. Many smartphones now have multiple sensors with different patterns, such as Quad-Bayer or Nona-Bayer. Most modern deep network-based models perform joint demosaicing and denoising with the strategy of training a separate network per pattern. Relying on individual models per pattern requires additional memory overhead and makes it challenging to switch quickly between cameras. In this work, we are interested in analyzing strategies for joint demosaicing and denoising for the three main mosaic layouts (1$\times$1 Single-Bayer, 2$\times$2 Quad-Bayer, and 3$\times$3 Nona-Bayer). We found concatenating a three-channel mosaic embedding to the input image and training a unified demosaicing architecture yields results that outperform existing Quad-Bayer and Nona-Bayer models and are comparable to Single-Bayer models. Additionally, we describe a maskout strategy that enhances the model performance and facilitates dead pixel correction---a step often overlooked by existing AI-based demosaicing models. As part of this effort, we captured a new demosaicing dataset of 638 RAW images that contain challenging scenes with patches annotated for training, validation, and testing. Code and data is available at \href{https://github.com/SamsungLabs/unified-demosaicing}{https://github.com/SamsungLabs/unified-demosaicing}.

\end{abstract}

}

\vspace{-2cm}
\ifpeerreview
\linenumbers \linenumbersep 15pt\relax 
\author{SaiKiran Tedla$^\texttt{1,2}$, Abhijith Punnappurath$^\texttt{1}$, Luxi Zhao$^\texttt{1}$, Michael S. Brown$^\texttt{1,2}$\\
$^\texttt{1}$ AI Center-Toronto, Samsung Electronics\hspace{0.4cm} $^\texttt{2}$ York University 
{\tt\small {\{s.tedla,abhijith.p,lucy.zhao,michael.b1\}@samsung.com}\hspace{0.4cm}\tt\small {\{tedlasai,mbrown\}@yorku.ca}}
}
\markboth{Anonymous ICCP 2025 submission ID \paperID}%
{}
\fi
\maketitle

\IEEEraisesectionheading{
  \section{Introduction and motivation}\label{sec:intro}
}
%
%
%
%

\begin{figure}[t]
  \centering
  \vspace*{-0.8cm}
  \includegraphics[width=1.00\linewidth]{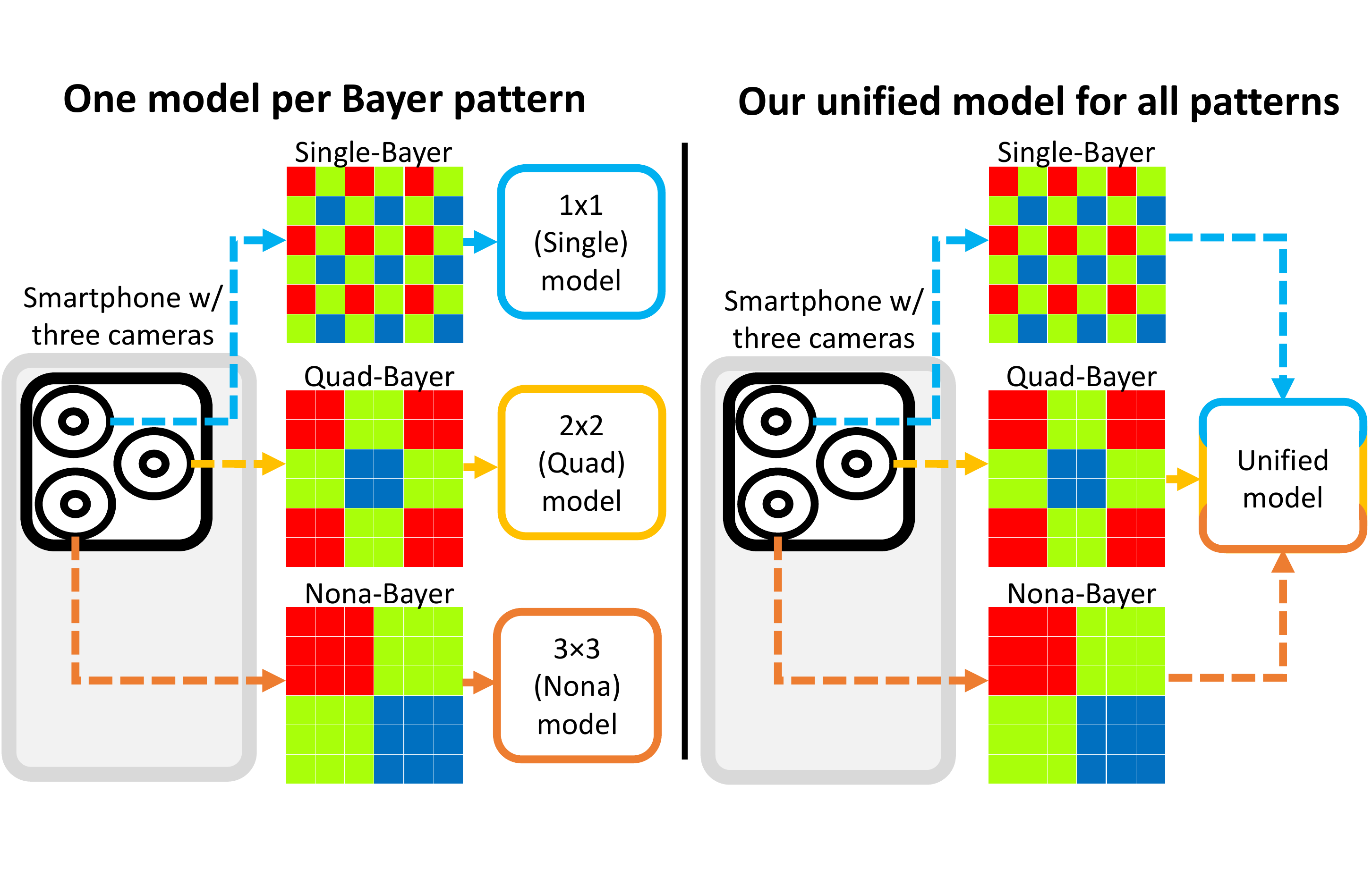}
  \vspace*{-1cm}
  \caption[Teaser]{Smartphones can now have multiple cameras with different patterns.  (Left) A current strategy is to use a separate demosaicing network for each pattern.  (Right) We explore unified models that demosaic all three common Bayer patterns. }\label{fig:teaser}
  \vspace*{-0.5cm}
  
\end{figure}

\IEEEPARstart{D}{emosaicing} is a key processing step applied by a camera's image signal processing (ISP) hardware~\cite{isp1} that estimates a full three-channel (RGB) image from the sensor's color filter mosaic layout. The standard Bayer pattern, we refer to as a Single-Bayer pattern, shown in Figure~\ref{fig:teaser}, was the dominant layout for many years. However, smartphones and devices are increasingly using new sensor patterns---namely Quad-Bayer~\cite{quadbayerexists} and Nona-Bayer~\cite{nonabayerexists}. Conventional demosaicing algorithms are unable to operate directly on these new patterns.  As a result, a common strategy is to rearrange or ``shuffle'' the layouts into a standard Single-Bayer arrangement and then process them by a Single-Bayer demosaicing algorithm. This process of rearranging the Quad-Bayer and Nona-Bayer patterns into a Single-Bayer pattern is known as remosaicing~\cite{remosaic1, remosaic2}. Naive remosaicing, such as shuffling, often results in poorly demosaiced images (we visualize in Section~\ref{sec:unified_results}).

  The current state-of-the-art in demosaicing uses deep network models, which are trained to perform joint demosaicing and denoising (e.g., ~\cite{bjdd, sagan,jdndm}). When deep networks are used, remosaicing is avoided by training a different model per pattern. While using individual models provides good results, it comes at the cost of additional memory overhead. Moreover, smartphones typically use multiple cameras to provide seamless zoom functionality~\cite{zoom}.  When users dynamically change their zoom factor, the ISP may be required to switch between two sensors with different CFA patterns. This rapid switching among sensors requires all demosaic models to be pre-loaded on the neural processing unit (NPU) or requires a noticeable delay when loading up a new model to the NPU.
  The impetus of this work is to find a {\it single} deep network model that handles different mosaic patterns as shown in Figure~\ref{fig:teaser}.
  
  \noindent{\textbf{Contribution}}~We propose an effective unified demosaicing/denoising model for Single-Bayer, Quad-Bayer, and Nona-Bayer patterns. We first examine the performance of deep network models targeting individual layouts at different noise levels to establish state-of-the-art baselines. Next, we examine three different strategies for designing a unified model. We show that a straightforward embedding of the pattern layout as part of the input provides the best results that are on par with individual models. We also demonstrate that this embedding approach lends itself well to a maskout augmentation strategy that not only improves performance but also naturally performs dead-pixel correction. Finally, as part of this effort, we produced a new challenging dataset for demosaicing containing 638 RAW images composed of high-frequency scene content.
  
  \section{Related work}
  \label{sec:background}


  \noindent{\textbf{Demosaicing.}} Early demosaicing algorithms centered around signal-processing-based strategies (e.g., bilateral and custom filters) often coupled with regularization based on spatio-spectral priors (e.g.,~\cite{model-based-1, gc1, model-based-3}).  Current state-of-the-art demosaicing methods are deep-learning-based and inspired by image-to-image translation tasks; common architectures such as CNNs~\cite{cnns, alexnet} and autoencoders~\cite{autoencoder} are used extensively.
  
  The first deep-learning-based demosaicing model~\cite{wang2014multilayer} used an autoencoder to learn demosaicing for $4\times4$ image patches. Then, others used a deep residual network with a two-stage approach~\cite{tan2017color} that first recovers the green channel and then estimates the red/blue channel information. Recently, models such as the dual pyramid network (DPN)~\cite{dpn} started targeting Quad-Bayer patterns.
  
  \noindent{\textbf{Joint demosaicing and denoising.}} Denoising is an important step of an ISP, because most consumer-grade image sensors exhibit noise~\cite{sidd}. Performing denoising initially may eliminate image detail that contributes to subsequent demosaicing steps. Conversly, performing demosaicing first can make it harder to denoise, because in the original mosaic the noise is decorrelated. One effective solution to this problem is to use a joint demosaicing and denoising model~\cite{jdndm}.
  
The first deep learning for joint demosaicing and denoising used a CNN based on the idea of residual prediction~\cite{paris}.  Additionally, this work selectively filtered the data, focusing specifically on ``hard patches'' for the demosaicing process. Then, others introduced an interative denoising network as a different approach~\cite{kokkinos2018deep}. Later works used green-channel self-guidance to direct the demosaicing process~\cite{sgnet}. Next, as large residual networks became widespread, JDNDM~\cite{jdndm} was released, a highly performant model built on the backbone of RCAN~\cite{rcan}. Other models such as BJDD~\cite{bjdd} and SAGAN~\cite{sagan} target joint demosaicing and denoising for Quad-Bayer and Nona-Bayer patterns, respectively. 

  Finally, other works may contain joint demosaicing and denoising within their architecture. For example, learning a low-light imaging pipeline can include joint demosaicing and denoising~\cite{seeinginthedark} . Others have also explored trinity nets~\cite{pixelshift200}: joint networks that handle demosaicing, denoising, and super-resolution. This paper will compare only against works that specifically address joint demosaicing and denoising.
  
  \noindent{\textbf{Remosaicing.}} Another approach for demosaicing new sensor patterns is remosaicing. Remosaicing models convert Quad-Bayer or Nona-Bayer mosaics into Single-Bayer mosaics, which are then passed to a Single-Bayer demosaicing network. The most basic type of remosaicing is shuffling pixels; we will show later that this is ineffective. Others build learning-based remosaicing models~\cite{remosaic1, remosaic2}. Quad-Bayer joint denoising and remosaicing was an ECCV 2022 challenge~\cite{remosaic_challenge}, and the winner was DRUNet~\cite{drunet}: a large network (112 MB) that combines U-Net~\cite{unet} and ResNet~\cite{resnet}. Any remosaicing algorithm must inherently output a one-channel mosaic. We will show this bottleneck limits remosaicing as a good solution for handling Quad-Bayer and Nona-Bayer patterns. Others~\cite{remosaic_doesnt_work} have suggested remosaicing may not be optimal because it can introduce artifacts.
  
  \noindent{\textbf{Unified-model for demosaicing.}} There is little work on building a demosaicing model that handles multiple pattern types. The only work in this area is KLAP~\cite{KLAP}; however, the authors have not made the training code or dataset public for this method. We re-implemented their training based on their descriptions. KLAP uses student-teacher learning and adaptive-discriminative filters to create a unified model. This work was not benchmarked against individual demosaicing methods, so it remains unclear if KLAP outperforms current individual state-of-the-art models like JDNDM, BJDD, and SAGAN. As mentioned in the introduction, the lack of focus on a unified model is the impetus of our work. 

  \section{Hard demosaicing dataset}
  \label{sec:dataset}

\begin{figure*}[h!]
    \centering
    \includegraphics[width=1.0\linewidth]{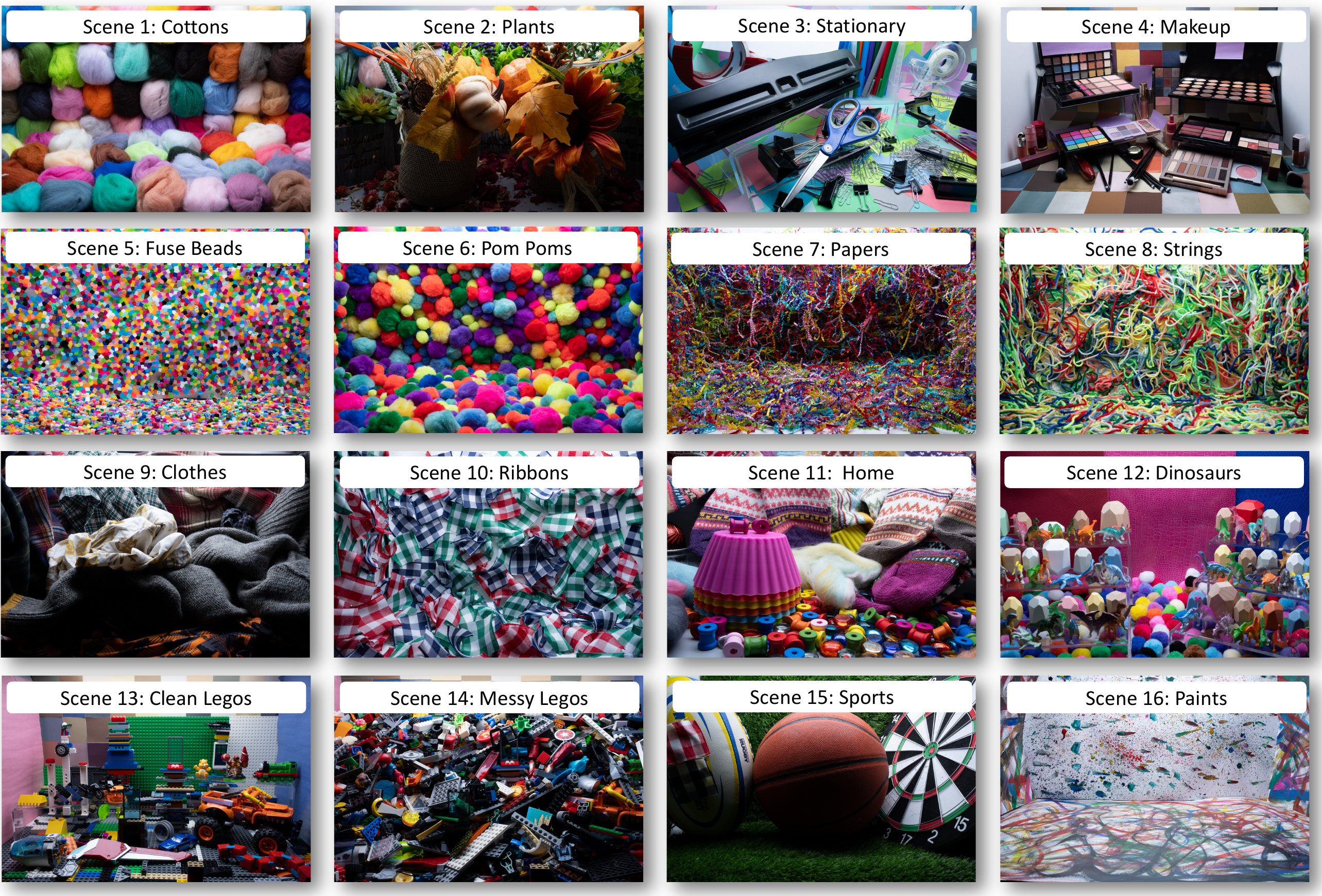}
    \caption[Scenes]{Our Hard Demosaicing Dataset consists of 638 RAW images captured from 17 scenes of different textured and small objects. We construct our scenes to contain more high-frequency detail that is challenging for demosaicing models.}\label{fig:scenes}
  \end{figure*}

  Before we examine demosaicing algorithms, we first describe the capture of our dataset, shown in Figure~\ref{fig:scenes}, as it will be used for training and evaluation of methods described in this paper.

  Our motivation to capture a new dataset is because many popular learning-based demosaicing methods (e.g.,~\cite{paris, jdndm, kokkinos2018deep, sagan}) were trained and evaluated on synthesized datasets.  These prior works synthesized Bayer inputs using processed sRGB images that had already had demosaicing applied.  Denoising and demosaicing are applied as early steps in the camera's ISP and operate on RAW sensor images.  As a result, we sought to create a dataset comprised of RAW images. 
  
  
  
  The only existing RAW image dataset for demosaicing~\cite{pixelshift200} contains 200 images that lack high-frequency details that are ``hard'' for demosaicing algorithms. To address these shortcomings, we capture our own RAW dataset of 638 images with carefully hand-crafted scenes that contain high-frequency detail. As a result, we name our dataset the ``Hard Demosaicing Dataset'' (HDD). We captured indoor scenes in a GTI lightbox~\cite{gti} with DC lighting to avoid flicker, brightness changes, and motion. We captured our dataset using a Sony DSLR camera with an FE 24-70mm GM II zoom lens at F/22 at ISO 100 in RAW. Finally, we capture our scenes in pixelshift mode to get ground truth RGB for each pixel.
  
HDD contains 638 images captured in 17 scenes shown in Figure~\ref{fig:scenes}. We construct each scene to contain high-frequency regions by placing textured or small objects.

  Each scene is captured from many views; all views vary in terms of position, orientation, and/or zoom. Each image captured in this dataset is $8640 \times 5760$ pixels. Due to the small pixel size on this sensor, we further downsample (by averaging each non-overlapping $2\times2$ pixel block) these images to $2160 \times 1440$ pixels to reduce noise. We observed that downsampling fixes the issue of slight out-of-focus blur that we encountered even when shooting at the smallest aperture setting (F/22). Additionally, we took special care to throw out images with aliasing artifacts or pixelshift artifacts (caused by movement between pixelshift captures). Finally, we check the noise of our clean images utilizing a single-image noise estimation method ~\cite{colom2013analysis} and found the highest bin mean across channels is 0.2099 (std: 0.0026), indicating near-zero noise. 
  

  \noindent{\textbf{Pre-processing}} Later, we will use HDD for testing demosaicing methods. We use a patch training framework~\cite{bjdd, sagan, jdndm} with a size of $48\times 48$ pixels. We used only the ``hardest'' 25\% of patches per view. We implemented a simple criterion for our hard-patch mining as we did not want to diminish our dataset's size heavily. First, we take the clean (ISO 100) ground-truth patches and mosaic them according to a Single-Bayer pattern. Then, we apply a bilinear interpolation as a simple demosaicing. Next, we sort the patches by reconstruction PSNR to find the hardest patches. We annotate the hard patch locations using the Single-Bayer configuration, then reuse this same set for Quad-Bayer and Nona-Bayer training. This design ensures that all CFA configurations are evaluated on the same hard patches, enabling a consistent and fair comparison in subsequent experiments. We split our train, validation, and test sets to contain scenes $1$--$10$ (113,045 hard patches), $11$--$12$ (11,470 hard patches), and $13$--$17$ (79,293 hard patches), respectively.
  
  Our training pairs consist of noisy mosaics and clean (ISO 100) full-channel RGB images. Following other demosaicing works~\cite{KLAP,pixelshift200}, we synthesize noisy mosaics from the clean images. We conducted experiments at four different ISO levels: 400, 800, 1600, and 3200. For each ISO level, we calibrate a Poisson-Gaussian model~\cite{plotz} for our sensor. We use 180 images of a color chart (30 images at 6 EVs) for calibration. We visualize our calbiration process and additional non-parametric noise models in our supplementary.

  \section{Individual models}
  \label{sec:individual_models}

  
  As mentioned in Section~\ref{sec:intro}, we will first explore individual models for demosaicing the different pattern types. Then, in Section~\ref{sec:unified_models} our unified models will be discussed.
  
  \noindent{\textbf{Existing individual demosaicing methods.}} We evaluated the current state-of-the-art individual demosaicing model for each pattern type. Thus, we use JDNDM for Single-Bayer, BJDD for Quad-Bayer, and SAGAN for Nona-Bayer. Additionally, we evaluated training JDNDM with shuffled Quad-Bayer and Nona-Bayer mosaics. We use shuffled data because JDNDM cannot be applied to Quad-Bayer and Nona-Bayer data directly because this network contains a ``packing'' convolution that splits the Single-Bayer image into a packed mosaic ($H/2 \times W/2 \times 4$). Training with shuffled data is inspired by naive shuffling as a remosaicing approach. The main difference is we trained with shuffled data, whereas shuffling for remosaicing is applied only at test time.
  
  \noindent{\textbf{Remosaicing methods.}}
  Remosaicing allows Quad-Bayer and Nona-Bayer data to be demosaiced with a pre-existing Single-Bayer demosaicing method. The most simple form of remosaicing is shuffling pixels such that they are arranged in a Single-Bayer pattern (see Figure~\ref{fig:model}). Remosaicing can also be learned given pairs of Quad-Bayer/Nona-Bayer mosaics and Single-Bayer mosaics. We can build these pairs using our dataset. In this framework, a Quad-Bayer/Nona-Bayer mosaic is passed into a remosaicing model and then to a Single-Bayer demosaicing model (in our case JDNDM). We will use DRUNet~\cite{drunet} for later experiments and show that learning remosaicing is not optimal.

  \subsection {Experiments}
  
  Our experiments use HDD and the splits mentioned in the Section~\ref{sec:dataset}. JDNDM is trained using the MSE loss. For BJDD and SAGAN, we use the original loss, a combination of a reconstruction (L1) loss, adversarial loss, and Delta E color loss (To be fair with other methods, we did try an MSE loss, but found the original loss to have higher PSNR). Additionally, we use a learning rate of $10^{-4}$ and a batch size of 16 for all experiments.
  
\begin{figure*}[h]
    \centering
    \includegraphics[width=1\linewidth]{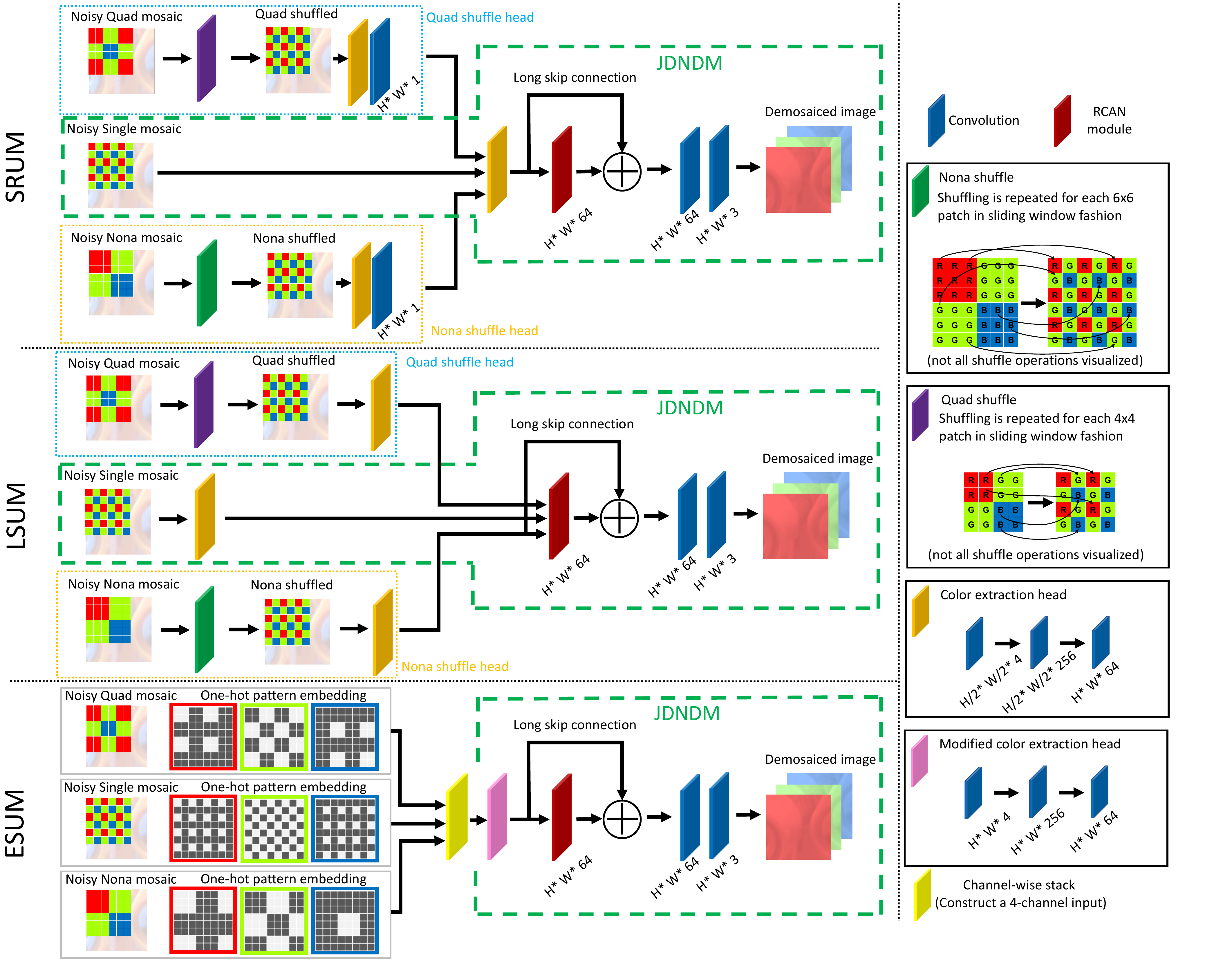}
    \caption[Model]{The three unified model approaches we compared. SRUM is a unified model based on remosaicing. LSUM is a unified model that uses a shared latent space and bypasses the bottleneck of remosaicing. For both SRUM and LSUM, we visualize the versions with shuffle heads; the packing head version will be visualized in supplementary. Finally, ESUM is our pattern embedding-based approach.}\label{fig:model}
  \end{figure*}

  Remosaicing experiments are tested by remosaicing Quad-Bayer or Nona-Bayer data into a Single-Bayer mosaic and then passing it through an existing Single-Bayer demosaicing network (in this case, the JDNDM we trained on Single-Bayer data). We use a deterministic shuffling process that is visualized in Figure~\ref{fig:model}. For learned remosaicing with DRUNet, we construct pairs of Quad-Bayer/Nona-Bayer to Single-Bayer mosaics with our HDD. We then train DRUNet with MSE loss to learn the remosaicing process. We report our reconstruction PSNR metric on the final image output by combining the remosaicing and demosaicing models.
  For all methods, we report reconstruction PSNR on hard patches in Table~\ref{tab:individual_models}.  PSNRs are reported after cropping a 2-pixel border around patches. We note high PSNRs are typical when working in RAW; see Table 1 of Brooks et al~\cite{brooks2019unprocessing}. Results for full images are in the supplementary material.

  \subsection {Discussion}
    We observed that training the JDNDM model with shuffled mosaics showed better performance compared to the BJDD and SAGAN models, which are currently recognized as the leading approaches for Quad-Bayer and Nona-Bayer patterns, respectively. This suggests that the necessity for specialized networks tailored to Quad-Bayer and Nona-Bayer patterns might not be as critical as previously thought. In addition, our observations highlight the versatility of the JDNDM model, as it consistently delivers good results across various scenarios. We will use this insight to construct our unified model in the Section~\ref{sec:unified_models}.
  
    We also noticed remosaicing by shuffling was extremely poor, which matches the intuition that shuffling diminishes spatial information. However, more interesting was that remosaicing with a large network like DRUNet (124.5 MB of parameters) did not perform as well as training JDNDM with shuffled data. This lower performance is probably because the remosaicing solution is forced to construct a one-channel mosaic before being passed into the Single-Bayer demosaicing network. Additionally, any remosaicing solution will inherently lose some information in the process of converting Quad-Bayer and Nona-Bayer mosaics into a Single-Bayer mosaic. The simple solution is to go directly from Quad-Bayer and Nona-Bayer mosaics to a full-channel RGB image, which is what we will do in our unified models.

  \label{sec:model}

  \section{Unified models}
  \label{sec:unified_models}
  
  We want to build a unified model that can handle Single-Bayer, Quad-Bayer, and Nona-Bayer approaches. To this end, we examined three different approaches. The first two are multi-headed approaches where each input head accepts a particular pattern type. The first multi-headed approach is inspired by remosaicing as we force all mosaics into a one-channel representation; we call this the standard remosaic unified model (SRUM). Our second multi-headed approach uses a unified latent space that circumvents the bottleneck implicit within remosaicing approaches: we call this model the latent-space unified model (LSUM). The last approach uses a one-hot pattern embedding to encode channel information. The fundamental similarity between all these approaches is that they are trained jointly on Single-Bayer, Quad-Bayer, and Nona-Bayer mosaics within a single iteration.  Figure~\ref{fig:model} illustrates all three approaches. All approaches are built off of the JDNDM architecture as we found this was the best individual model for all patterns.

\begin{table*}[ht]

    \caption[]{Results of different individual demosaicing methods on all three pattern types. We report results for BJDD~\cite{bjdd}, SAGAN~\cite{sagan}, and JDNDM~\cite{jdndm} on the pattern type they were designed for. We also report results for a version of JDNDM trained with shuffled Quad-Bayer and Nona-Bayer data. For remosaicing approaches (shuffle and DRUNet~\cite{drunet}), JDNDM is used as the Single-Bayer demosaicing network, and we report the size of the remosaicing model plus the size of JDNDM. In shuffle remosaicing, there is no model, so we only report the size of JDNDM. Results are reported at all 4 ISO levels. Green, yellow, and blue highlighting represents first, second, and third best, respectively.
    }
    \resizebox{1.0\textwidth}{!}{%
        \begin{tabular}{@{}lccccccccccccc}
            \toprule \multicolumn{1}{c}{}                         & \multicolumn{1}{c|}{} & \multicolumn{12}{c}{\textbf{Hard patches (PSNR)$\uparrow$}}                                                                                                                                                                                                                                                                                                                                                                 \\
            \cline{3-14}
            \multicolumn{1}{c}{}                     & \multicolumn{1}{c|}{\multirow{2}{1.3cm}{\centering \textbf{ Size (MB) $\downarrow$ } }}                                                                        & \multicolumn{3}{c|}{\textbf{ISO 400}}                       & \multicolumn{3}{c|}{\textbf{ISO 800 }} & \multicolumn{3}{c|}{\textbf{ISO 1600}} & \multicolumn{3}{c}{\textbf{ISO 3200}}                                                                                                                                                                                                                                       \\

            \multicolumn{1}{l}{\multirow{-1}{*}{\textbf{Method}}} & \multicolumn{1}{c|}{}                                                                           & \multicolumn{1}{c}{Single}                                  & Quad                                   & \multicolumn{1}{c|}{Nona}              & Single                                & Quad                       & \multicolumn{1}{c|}{Nona } & Single                    & Quad                       & \multicolumn{1}{c|}{Nona}  & Single                    & Quad                       & Nona                       \\
            \toprule

            BJDD                                                  & 13.29                                                                                           & -                                                           & \cellcolor{blue!30}50.86                & -                                      & -                                     & \cellcolor{blue!30} 50.05                      & -                          & -                         & \cellcolor{blue!30}48.88    & -                          & -                         & 47.50                      & -                          \\

            SAGAN                                                 & 112.34                                                                                          & -                                                           & -                                      & \cellcolor{blue!30} 49.55                                  & -                                     & -                          & \cellcolor{blue!30} 49.06                      & -                         & -                          & \cellcolor{blue!30}48.05    & -                         & -                          & \cellcolor{blue!30}46.88    \\

            JDNDM                                                 & 12.21                                                                                           & \cellcolor{green!30}53.69                                   & -                                      & -                                      & \cellcolor{green!30}52.34             & -                          & -                          & \cellcolor{green!30}50.43 & -                          & -                          & \cellcolor{green!30}49.00 & -                          & -                          \\

            JDNDM  - Shuffle                                      & 12.21                                                                                          & -                                                           & \cellcolor{green!30}52.15              & \cellcolor{green!30}51.32              & -                                     & \cellcolor{green!30}51.11  & \cellcolor{green!30}50.30  & -                         & \cellcolor{green!30}49.82  & \cellcolor{green!30}49.23  & -                         & \cellcolor{green!30}48.46  & \cellcolor{green!30}47.86  \\

            Remosaic Shuffle                                      & 12.21                                                                                          & -                                                           & 40.42                                  & 36.48                                  & -                                     & 40.55                      & 36.60                      & -                         & 40.62                      & 36.73                      & -                         & 40.67                      & 36.85                      \\

            Remosaic DRUNet                                       & 148.81                                                                                         & -                                                           & \cellcolor{yellow!35}51.78             & \cellcolor{yellow!35}51.00             & -                                     & \cellcolor{yellow!35}50.57 & \cellcolor{yellow!35}49.93 & -                         & \cellcolor{yellow!35}49.31 & \cellcolor{yellow!35}48.86 & -                         & \cellcolor{yellow!35}48.02 & \cellcolor{yellow!35}47.71 \\

            \bottomrule

        \end{tabular}
    }

    \label{tab:individual_models}
\end{table*}

  \begin{table*}[t]
\small
\centering
\caption{Results of our three types of unified models and KLAP~\cite{KLAP}. For SRUM and LSUM, we try models with both shuffle heads and packing heads. PSNR is reported on hard patches and full images. FLOPs are always reported on $256\times256$ patches. Green, yellow, and blue highlighting represents first, second, and third best, respectively.}
\resizebox{1.0\textwidth}{!}{%
\begin{tabular}{@{}lcc|cccccccccccc@{}}
				\toprule \multicolumn{1}{c}{}                         & \multicolumn{12}{c}{\textbf{Hard patches (PSNR)$\uparrow$}}                                                                                                                                                                                                                                                                                                                                                                                                                                               \\
\toprule
\multicolumn{1}{c}{} &  && \multicolumn{3}{c|}{\textbf{ISO 400}} & \multicolumn{3}{c|}{\textbf{ISO 800}} & \multicolumn{3}{c|}{\textbf{ISO 1600}} & \multicolumn{3}{c}{\textbf{ISO 3200}} \\
\cmidrule(lr){4-6} \cmidrule(lr){7-9} \cmidrule(lr){10-12} \cmidrule(lr){13-15}
\textbf{Method} &  \multicolumn{1}{c|}{\textbf{Size (MB) $\downarrow$}}  &  \multicolumn{1}{c|}{\textbf{FLOPs (G) $\downarrow$}}  & Single & Quad & Nona & Single & Quad & Nona & Single & Quad & Nona & Single & Quad & Nona \\
\toprule
KLAP & 25.62 & 106 & 53.27 & 52.01 & 50.44 & 51.91 & 50.99 & 49.79 & 50.28 & 49.40 & 48.59 & \cellcolor{blue!30}48.95 & 48.10 & 47.20 \\
SRUM Shuffle & 14.01 & 415 & 53.33 & 51.83 & 50.92 & \cellcolor{blue!30}52.10 & 50.89 & 50.24 & 50.35 & 49.46 & 48.99 & 48.91 & 48.19 & 47.76 \\
SRUM Packing & 16.53 & 415 & 53.12 & 51.52 & 50.58 & 51.70 & 50.39 & 49.70 & \cellcolor{blue!30}50.36 & 49.16 & 48.82 & 48.87 & 48.25 & 47.71 \\
LSUM Shuffle & 13.41 & 415 & \cellcolor{blue!30}53.47 & \cellcolor{yellow!35}52.38 & \cellcolor{yellow!35}51.58 & \cellcolor{green!25}52.17 & \cellcolor{blue!30}51.16 & \cellcolor{yellow!35}50.63 & 50.28 & \cellcolor{yellow!35}49.84 & \cellcolor{yellow!35}49.13 & \cellcolor{green!25}49.03 & \cellcolor{green!25}48.58 & \cellcolor{yellow!35}48.09 \\
LSUM Packing & 15.92 & 415 & \cellcolor{yellow!35}53.55 & \cellcolor{blue!30}52.27 & \cellcolor{blue!30}51.37 & 52.08 & \cellcolor{yellow!35}51.22 & \cellcolor{blue!30}50.42 & \cellcolor{yellow!35}50.50 & \cellcolor{blue!30}49.71 & \cellcolor{blue!30}49.03 & 48.93 & \cellcolor{blue!30}48.38 & \cellcolor{blue!30}47.77 \\
ESUM & 12.21 & 408 & \cellcolor{green!25}53.75 & \cellcolor{green!25}52.68 & \cellcolor{green!25}51.96 & 52.17\cellcolor{green!25} & 51.36\cellcolor{green!25} & 50.76\cellcolor{green!25} & \cellcolor{green!25}50.64 & \cellcolor{green!25}50.01 & \cellcolor{green!25}49.46 & \cellcolor{yellow!35}48.98 & \cellcolor{yellow!35}48.57 & \cellcolor{green!25}48.11 \\
\bottomrule
\end{tabular}
}
\label{tab:unified_models}
\end{table*}

\begin{table}[t]

    \small
    \centering
    \caption{Results on Pixelshift200~\cite{pixelshift200}. We compare individual models (JDNDM and JDNDM-Shuffle) and KLAP~\cite{KLAP} with our unified model, ESUM. Green, yellow, and blue highlighting represents first, second, and third best, respectively.}
    
    \resizebox{0.45\textwidth}{!}{%
        \begin{tabular}{l>{\centering\arraybackslash}cccc}
            \toprule
            \multicolumn{1}{c}{}       & & \multicolumn{3}{c}{\textbf{Pixelshift200 (PSNR)$\uparrow$}}                                                                  \\

            \cline{3-5} \textbf{Method}        & \textbf{Size (MB) $\downarrow$}                                                               & Single                                  & Quad& Nona\\
            \toprule
            JDNDM        & 12.21                                                                                  & \cellcolor{yellow!35}55.87                                          & -                               & -                            \\

            JDNDM-Shuffle & 12.21                                                                                   & -                                                                   & \cellcolor{yellow!35}55.84      & \cellcolor{yellow!35}55.54   \\

            KLAP                       & 25.62                                                                                   & \cellcolor{blue!30}55.79                                             & \cellcolor{blue!30}55.64         & \cellcolor{blue!30}55.40
            \\
            ESUM                   & 12.21                                                                                   & \cellcolor{green!25}56.07                                           & \cellcolor{green!25}56.04       & \cellcolor{green!25}55.91    \\
        \end{tabular}
    }

    \label{tab:pixelshift200}
\end{table}

    \begin{figure*}[h!]
        \centering
        \includegraphics[width=0.95\linewidth]{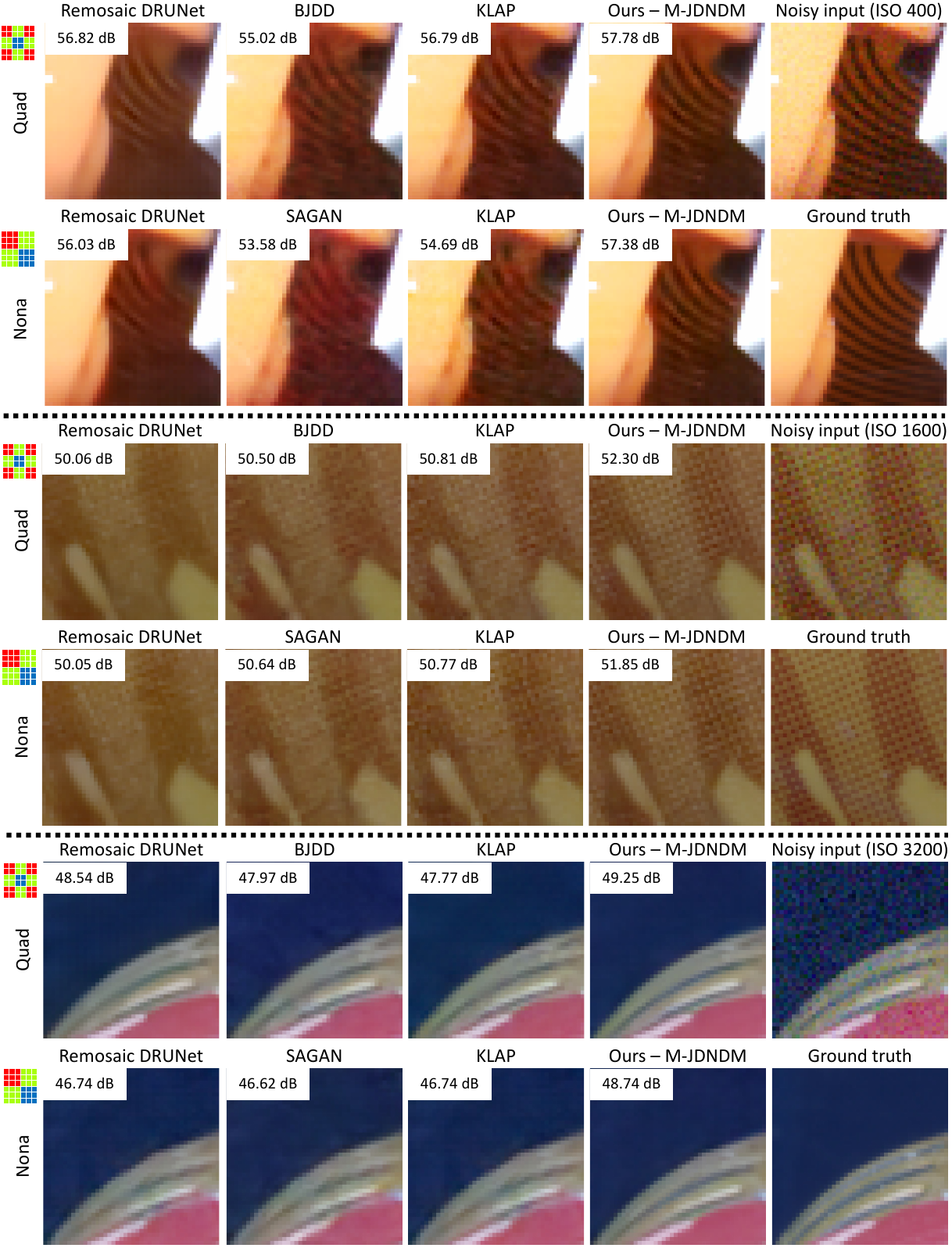}
        \caption[Qualitative]{Qualitative results comparing our embedding-based unified model, ESUM, with existing individual demosaicing methods (remosaic with DRUNet~\cite{drunet}, BJDD~\cite{bjdd}, SAGAN~\cite{sagan}) and a unified method, KLAP~\cite{KLAP}, for Quad-Bayer and Nona-Bayer mosaics. We show patches at ISO 400, 1600, and 3200 (noisy input is before mosaic sampling). PSNR is reported in RAW, but visualized images are rendered by an ISP~\cite{sidd}.} \label{fig:examples}
      \end{figure*}
  
  \noindent{\textbf{Standard remosaic unified model (SRUM).}} Demosaicing on all three pattern types is inherently similar; we propose a multi-head architecture where each head accepts a different pattern type, and a shared backbone is used. Our first approach with a multi-headed architecture builds on the principle of remosaicing. We use lightweight Quad-Bayer and Nona-Bayer heads that are passed as input into a Single-Bayer demosaicing network JDNDM. The main difference between SRUM and traditional remosaicing is that the remosaicing and demosaicing networks are trained jointly.
  
  For our experiments, we tested SRUM with two variations of remosaicing heads. The first variation is a shuffle head that shuffles the Quad-Bayer and Nona-Bayer data into a Single-Bayer mosaic and uses the same packing and unpacking convolutions found in JDNDM. SRUM with this packing head variation is visualized in Figure~\ref{fig:model}. The second variation is an approach that respects the Quad-Bayer and Nona-Bayer position information by changing the packing and unpacking convolutions. This results in packing the Quad-Bayer mosaic into $H/4 \times W/4 \times 16$ and the Nona-Bayer mosaic into $H/6 \times W/6 \times 36$. SRUM with a packing head is visualized in the supplementary material.

  \noindent{\textbf{Latent-space unified model (LSUM).}}
  Our unified approach based on remosaicing (SRUM) is naturally bottlenecked because it must convert to a single-channel mosaic with dimension $H \times W \times 1$.
  We examine if removing this bottleneck could improve performance. We do this by instead converting each mosaic into a shared latent space of dimension $H \times W \times 64$. Thus, LSUM is built by taking SRUM and removing the convolution that forces the Quad-Bayer and Nona-Bayer heads into a single channel. Another key difference is that the combined space is before the RCAN module. Hence, we can think about the model as having three separate heads: one for each pattern. We test the same variations of LSUM that we discussed for SRUM: one with shuffle heads and one with packing heads.
  
  \noindent{\textbf{Embedding-supervised unified model (ESUM).}} Our last approach, ESUM, is a modified version of JDNDM that handles multiple patterns with a pattern embedding. ESUM takes a four‑channel tensor
$\mathbf{x}\in\mathbb{R}^{H\times W\times 4}$.
The first channel, $\mathbf{x}_{1}$, contains the raw
mosaic intensities. Channels $\mathbf{x}_{2}$–$\mathbf{x}_{4}$ provide a
per-pixel one-hot encoding of a CFA defined as
\begingroup
\[
(\mathbf{x}_{2},\mathbf{x}_{3},\mathbf{x}_{4})(i,j)=
\begin{cases}
(1,0,0), & \text{if pixel $(i,j)$ has \textbf{red} filter},\\[4pt]
(0,1,0), & \text{if pixel $(i,j)$ has \textbf{green} filter},\\[4pt]
(0,0,1), & \text{if pixel $(i,j)$ has \textbf{blue} filter}.\\
\end{cases}
\]
\endgroup
Since the masks are aligned with the mosaic grid, they encode the CFA layout (Single-, Quad-, or Nona=Bayer). This mask gives the network explicit spatial knowledge of which
color filter produced each pixel value.

Another difference is that we replace the color extraction head of JDNDM with a color extraction head that does not use a packing convolution. Figure~\ref{fig:model} shows our modified network. Removing the packing convolution also removes a natural encoding of pattern information, and this is why we introduced the one-hot pattern embedding.

  \subsection{Experiments}
  We use a joint training approach for all our unified models where each iteration learns demosaicing for Single-Bayer, Quad-Bayer, and Nona-Bayer mosaics. All hyperparameters are identical to the individual model experiments. We build data with the same process described in the Section~\ref{sec:dataset}, but now we create a noisy mosaic for all three pattern types. Since we use a batch size of 16, this results in 48 mosaics constructed per batch. For SRUM and LSUM, we split mosaics through the heads depending on the pattern type of the mosaic. Then, once the mosaics pass through the heads, the intermediate representations for all pattern types are passed through the shared backbone. ESUM concatenates a one-hot pattern embedding to each of the 48 mosaics and trains the network with these four-channel inputs. Finally, we compare these approaches against KLAP~\cite{KLAP} which is retrained on our HDD dataset.

  \subsection{Results}
  \label{sec:unified_results}

  Table~\ref{tab:unified_models} contains the results of our different unified model approaches. ESUM outperformed other approaches under most settings. ESUM shows a one-hot pattern embedding is enough for the network to handle all three pattern types. Tables~\ref{tab:individual_models} and~\ref{tab:unified_models} show that ESUM outperforms BJDD and SAGAN on Quad-Bayer and Nona-Bayer demosaicing and matches the same performance of JDNDM on Single-Bayer demosaicing. We also find ESUM outperforms baselines in the full-image case and on other metrics such as SSIM and Delta E (see supplementary for results). Figure~\ref{fig:examples} shows qualitative results for Quad-Bayer and Nona-Bayer demosaicing; we visualize these patterns as they are new and have created the need for a unified model. This visualization compares ESUM with a mosaic maskout augmentation (discussed in the Section~\ref{sec:maskout}) against existing individual and unified methods. Additional qualitative results at all four ISO levels and are visualized in the supplementary. 
  

      ESUM outperforms SRUM because SRUM bottlenecks Quad-Bayer and Nona-Bayer mosaics by forcing remosaicing to a single-channel representation. LSUM works better than SRUM because it removes this bottleneck and creates a shared latent space for all patterns. The higher performance of ESUM could be because ESUM can use the full network capacity to interpret spatial information, while LSUM can only do this in the heads. Additionally, our methods follow the same paradigm of jointly performing demosaicing and denoising~\cite{jdndm, KLAP}, however, SRUM/LSUM alter inter-pixel distances via shuffling, potentially harming denoising. In contrast, ESUM preserves true inter-pixel distances and leverages pattern-embedded convolutions for improved denoising. Additional qualitative results comparing ESUM with SRUM and LSUM are in the supplementary material.

      Finally, we note that ESUM outperforms the baseline KLAP in terms of reconstruction metrics (e.g., +1.5dB on ISO 400, Nona-Bayer). We note that models (SRUM, LSUM, and ESUM) based on the JDNDM architecture have less parameters but more FLOPs than the baseline KLAP. However, we observe that the runtime (measured on a RTX 6000 for a $256\times256$ patch) is 15 ms for ESUM and 18 ms for KLAP. We believe our model could be further improved through optimization of the various channel attention and convolution mechansims~\cite{nafnet}. 

\subsection{Pixelshift200}

Additionally, we test our best unified model, ESUM, on the Pixelshift200~\cite{pixelshift200} dataset. Pixelshift200 contains RAW images for demosaicing but these images contain less high-frequency detail than our dataset. We used 200,000 patches ($48\times48$) from Pixelshift200 and evaluated with a 70/10/20 split. We use the original noise distribution from the dataset and find it is similar to the ISO 3200 noise in our HDD; we observe this qualitatively and also quantitatively as the shot/read noise parameters are within $\pm 0.0001$. We compare ESUM against KLAP~\cite{KLAP} and the best individual models (JDNDM and JDNDM - Shuffle). We report results in Table~\ref{tab:pixelshift200}.

We found ESUM outperforms KLAP and the best individual models on Pixelshift200. This result shows our approach generalizes to other datasets. By comparing Table~\ref{tab:unified_models} and Table~\ref{tab:pixelshift200} we notice that ESUM is on average 7.5 dB higher on Pixelshift200 patches compared to our ISO 3200 HDD patches. As mentioned before, the noise levels between Pixelshift200 and ISO 3200 HDD are similar, so this 7.5 dB gap can be explained by the lack of high-frequency content in Pixelshift200. This is further confirmed by noticing the PSNR difference of ESUM between Single-Bayer and Nona-Bayer is only 0.16 dB. On our ISO 3200 HDD, this gap is 0.87 dB. Both observations are consistent with our motivation to create a challenging dataset for demosaicing.

\subsection{Demosaicing Only}

We test our best unified model, ESUM, on the demosaicing only scenario (no noise) on our HDD dataset. We compare ESUM against KLAP. Additionally, for the Single-Bayer scenario we compare with some standard non-learning methods including bilinear interpolation, OpenCV Edge-Aware~\cite{bradski2000opencv}, and Menon~\cite{menon2006demosaicing}. As seen in Table~\ref{tab:demosaicingonly}, we find ESUM to outperform all methods in this scenario.

\begin{table}[t]

    \small
    \centering
    \caption{Results on demosaicing only scenario. We compare non-learning methods (Bilinear, Edge-Aware~\cite{bradski2000opencv}, and Menon~\cite{menon2006demosaicing}) and KLAP~\cite{KLAP} with our unified model, ESUM. Green, yellow, and blue highlighting represents first, second, and third best, respectively.}
    
    \resizebox{0.45\textwidth}{!}{%
        \begin{tabular}{l>{\centering\arraybackslash}cccc}
            \toprule
            \multicolumn{1}{c}{}       & & \multicolumn{3}{c}{\textbf{Hard patches (PSNR)$\uparrow$}}                                                                  \\

            \cline{3-5} \textbf{Method}        & \textbf{Size (MB) $\downarrow$}                                                               & Single                                  & Quad& Nona\\
            \toprule
            Bilinear        & 0.00                                                                                 & 46.50                                          & -                               & -                            \\

            Edge-Aware & 0.00                                                                                  & 45.23                                                                  & -      & -  \\
            Menon & 0.00                                                                                  & \cellcolor{blue!30}46.91                                                                   & -     &  - \\
            KLAP                       & 25.62                                                                                   & \cellcolor{yellow!35}58.25                                             & \cellcolor{yellow!35}55.73        & \cellcolor{yellow!35}54.02
            \\
            ESUM                   & 12.21                                                                                   & \cellcolor{green!25}58.98                                          & \cellcolor{green!25}56.31       & \cellcolor{green!25}55.30    \\
        \end{tabular}
    }

    \label{tab:demosaicingonly}
\end{table}

\subsection{Binning Modes}
\begin{figure}[h!]
    \centering
    \includegraphics[width=1\linewidth]{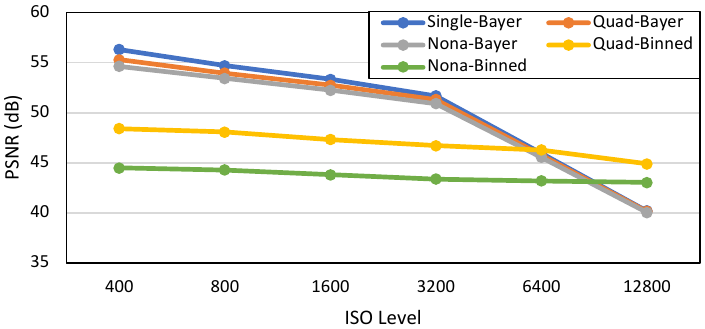}
    \caption[Binning]{Tradeoffs of standard and binned modes when demosaicing. We visualize at each ISO the results of demosaicing with Single-Bayer, Quad-Bayer, Nona-Bayer, Quad-Binned, and Nona-Binned. Quad-Binned and Nona-Binned use bicubic upsampling after demosaicing. We use ESUM trained at the respective ISO level for generating these results. }\label{fig:binning}
  \end{figure}

We evaluate our best unified model, ESUM, on the task of demosaicing under binning modes, a common sensor technique that combines multiple pixels into one to improve noise performance in low-light conditions. Specifically, we simulate binning by summing non-overlapping $2\times2$ patches in Quad-Bayer and $3\times3$ patches in Nona-Bayer mosaics, effectively reducing resolution by factors of 2 and 3, respectively. This process yields a Single-Bayer binned mosaic to which simulated noise is added. The noisy binned mosaics are demosaiced and denoised using ESUM in Single-Bayer mode. Finally, we use bicubic upsampling to restore the original resolution. To simulate higher noise RAW images (low-light conditions), we additionally add noise levels at ISO 6400 and ISO 12800. We visualize the results of demosaicing at various ISO levels with and without binning modes in Figure~\ref{fig:binning}.

These results highlight the benefits of binning in high-noise settings and illustrate tradeoffs between binning strategies and CFA types as noise levels vary. Single-Bayer mosaics are the best in low-noise scenarios and allow for high-frequency detail to be recovered, wheras Quad-Bayer and Nona-Bayer provided flexible alternatives that allow for some high-frequency detail recovery (at low ISOs), but also provide better noise performance at higher ISOs.

      \section{Maskout augmentation/dead pixel correction}
      \label{sec:maskout}

      \begin{table*}[h]
    \small
    \centering
    \caption{Results showing how our maskout augmentation improves performance (Top) or improves performance when 1\% of pixels are simulated as dead (Bottom). We compare no maskout and two levels of maskout (0\%-1\% and 0\%-5\%). The no maskout model uses a $7\times7$ Gaussian filter for dead pixel interpolation; this is comparable to a standard ISP. Green, yellow, and blue highlighting represents first, second, and third best, respectively. }

    \resizebox{1.0\textwidth}{!}{%
        \renewcommand{\multirowsetup}{\centering}
        \begin{tabular}{l>{\centering\arraybackslash}cccccccccccc}
            \toprule
            \multicolumn{1}{c|}{}               & \multicolumn{12}{c}{\textbf{Maskout performance improvement (PSNR) $\uparrow$ }}                                                                                                                                                                                                                                  \\
            \cline{2-13}  \multicolumn{1}{c|}{} & \multicolumn{3}{c|}{\textbf{ISO 400}}                                                            & \multicolumn{3}{c|}{\textbf{ISO 800}} & \multicolumn{3}{c|}{\textbf{ISO 1600}} & \multicolumn{3}{c}{\textbf{ISO 3200}}                                                                                                         \\
            \multicolumn{1}{l|}{Maskout range}  & {Single  }                                                                                       & Quad                                  & \multicolumn{1}{c|}{Nona }             & Single                                & Quad & \multicolumn{1}{c|}{Nona } & Single & Quad & \multicolumn{1}{c|}{Nona } & Single & Quad & Nona \\
            \toprule
            No maskout                          & \cellcolor{green!25}53.75                                                                        & \cellcolor{green!25}52.68             & \cellcolor{yellow!35}51.96
                                                & \cellcolor{blue!30}52.17                                                                          & \cellcolor{blue!30}51.36               & \cellcolor{blue!30}50.76
                                                & \cellcolor{yellow!35}50.64                                                                       & \cellcolor{yellow!35}50.01            & \cellcolor{blue!30}49.46
                                                & \cellcolor{blue!30}48.98                                                                          & \cellcolor{blue!30}48.57               & \cellcolor{blue!30}48.11
            \\
            0\%~- 1\%                           & \cellcolor{blue!30}53.55                                                                          & \cellcolor{blue!30}52.51               & \cellcolor{blue!30}51.75
                                                & \cellcolor{green!25}52.38                                                                        & \cellcolor{green!25}51.66             & \cellcolor{green!25}51.06
                                                & \cellcolor{blue!30}50.53                                                                          & \cellcolor{blue!30}49.99               & \cellcolor{blue!30}49.46
                                                & \cellcolor{yellow!35}49.06                                                                       & \cellcolor{yellow!35}48.65            & \cellcolor{yellow!35}48.19
            \\
            0\%~- 5\%                           & \cellcolor{yellow!35}53.73                                                                       & \cellcolor{green!25}52.68             & \cellcolor{green!25}51.97
                                                & \cellcolor{yellow!35}52.30                                                                       & \cellcolor{yellow!35}51.50            & \cellcolor{yellow!35}50.85
                                                & \cellcolor{green!25}50.78                                                                        & \cellcolor{green!25}50.16             & \cellcolor{green!25}49.65
                                                & \cellcolor{green!25}49.10                                                                        & \cellcolor{green!25}48.69             & \cellcolor{green!25}48.26                                                                                                                                                              \\
                            
                                                \toprule
                                                \multicolumn{1}{c|}{}               & \multicolumn{12}{c}{\textbf{Maskout performance with dead pixels (PSNR) $\uparrow$}}                                                                                                                                                                                                                                  \\
                                                \cline{2-13}  \multicolumn{1}{c|}{} & \multicolumn{3}{c|}{\textbf{ISO 400}}                                                                 & \multicolumn{3}{c|}{\textbf{ISO 800}} & \multicolumn{3}{c|}{\textbf{ISO 1600}} & \multicolumn{3}{c}{\textbf{ISO 3200}}                                                                                                         \\
                                                \multicolumn{1}{l|}{Maskout range}  & {Single  }                                                                                            & Quad                                  & \multicolumn{1}{c|}{Nona }             & Single                                & Quad & \multicolumn{1}{c|}{Nona } & Single & Quad & \multicolumn{1}{c|}{Nona } & Single & Quad & Nona \\
                                                \toprule
                                                No maskout                          & \cellcolor{blue!30}52.65                                                                               & \cellcolor{blue!30}51.80               & \cellcolor{blue!30}51.09
                                                                                    & \cellcolor{blue!30}51.41                                                                               & \cellcolor{blue!30}50.74               & \cellcolor{blue!30}50.14
                                                                                    & \cellcolor{blue!30}50.18                                                                               & \cellcolor{blue!30}49.62               & \cellcolor{blue!30}49.07
                                                                                    & \cellcolor{blue!30}48.72                                                                               & \cellcolor{blue!30}48.35               & \cellcolor{blue!30}47.88
                                    
                                                \\
                                                0\%~- 1\%                           & \cellcolor{yellow!35}53.34                                                                            & \cellcolor{yellow!35}52.30            & \cellcolor{yellow!35}51.52
                                                                                    & \cellcolor{green!25}52.27                                                                             & \cellcolor{green!25}51.54             & \cellcolor{green!25}50.92
                                                                                    & \cellcolor{yellow!35}50.41                                                                            & \cellcolor{yellow!35}49.85            & \cellcolor{yellow!35}49.30
                                                                                    & \cellcolor{yellow!35}48.99                                                                            & \cellcolor{yellow!35}48.56            & \cellcolor{yellow!35}48.09
                                    
                                                \\
                                                0\%~- 5\%                           & \cellcolor{green!25}53.64                                                                             & \cellcolor{green!25}52.59             & \cellcolor{green!25}51.88
                                                                                    & \cellcolor{yellow!35}52.23                                                                            & \cellcolor{yellow!35}51.44            & \cellcolor{yellow!35}50.79
                                                                                    & \cellcolor{green!25}50.72                                                                             & \cellcolor{green!25}50.11             & \cellcolor{green!25}49.60
                                                                                    & \cellcolor{green!25}49.05                                                                             & \cellcolor{green!25}48.64             & \cellcolor{green!25}48.21                                                                                                                                                              \\
        \end{tabular}
    }

    \label{tab:maskout}
\end{table*}

      Our embedding-based unified model, ESUM, shows good performance when demosaicing all three pattern types. However, we found a simple augmentation which we call ``mosaic maskout'' can improve the performance of ESUM. Mosaic maskout works by removing (setting to zero) some pixels from the mosaic at training time; this is a regularization for demosaicing. Our augmentation is viable because of the pattern-based embedding that can be updated to reflect which pixels are dropped from the mosaic.
      
      We implement mosaic maskout by randomly dropping some pixels from each mosaic image in any training batch. For our implementation, we try two levels of maskout. One level randomly samples 0\%--1\% of pixels for maskout; the other level samples 0\%--5\% of pixels for maskout. We randomly sample pixels between this type of interval so the network sees mosaics with varying pixels dropped out; additionally, we still want the model to work well when no pixels are dropped. Finally, we update the one-hot pattern embedding to indicate the dropped pixels. Quantitive performance improvements are given in Table~\ref{tab:maskout} (top). 

      \subsection{Dead pixel correction}
      
      Mosaic maskout is not only useful as an augmentation for improving performance but also is important because sensors have ``dead'' pixels that must be corrected~\cite{gc1}. A dead pixel mask is typically calibrated by the sensor manufacturer and can account for as much as 1\% of the total number of pixels~\cite{sidd}. Typically, dead pixels are interpolated before demosaicing by replacing the dead pixel value with a weighted average of surrounding pixels with the same color channel~\cite{deadpixel1}. However, this interpolation is imperfect and can lead to artifacts in the final image. Additionally, interpolating across these dead pixels removes information that could aid the demosaicing process.
      
      For our baseline (no maskout), we use traditional interpolation to replace dead pixels. We implement the traditional interpolation baseline by taking a weighted average using a $7 \times 7$ Gaussian filter with $\sigma$=$3$; this filter weights only pixels that match the color channel of the dead pixel. For our models trained with maskout, we set dead pixels to 0 and updated the pattern mask accordingly. We compare all models at a dead pixel rate of 1\%; results are in Table~\ref{tab:maskout} (bottom). Our model with mosaic maskout outperforms the tradtiional interpolation baseline and can demosaic with dead pixels more accurately. We also notice that using a larger range for maskout (0\%--5\%) improved results better than the 0\%--1\%, showing it can be useful to train with maskout ranges that have patches larger than the dead pixel rate of the sensor (1\% in this case). Additionally, utilizing our maskout augmentation removes the need for a separate dead pixel interpolation step on the ISP.

      

      \section{Summary}
       
        This paper has shown that our embedding-supervised unified demosaicing model, ESUM, can outperform current state-of-the-art individual demosaicing models for Quad-Bayer and Nona-Bayer patterns while matching performance on Single-Bayer patterns. Our model can effectively demosaic multiple pattern types \textit{and} it is the first for joint demosaicing, denoising, and dead-pixel correction.
      
        We have also shown that remosaicing approaches and pattern-specific architectures are not required for demosaicing the new Quad-Bayer and Nona-Bayer patterns. We explored three different approaches for a unified model and showed our embedding-based solution, ESUM, was better than other approaches on our HDD dataset and Pixelshift200~\cite{pixelshift200}. Additionally, we have shown that a mosaic maskout augmentation can improve the performance of ESUM and be used to correct dead pixel readings from the sensor. Our approach is simple, effective, and appealing to camera manufacturers as it combines three steps of the ISP.
      
        Finally, we produced HDD, a RAW demosaicing dataset which contains challenging images for testing demosaicing applications. 

          \noindent{\textbf{Limitations.}} A main limitation of our work is that our model is fixed per ISO level and does not adapt to different noise levels.  A true heterogeneous system is much more challenging. For example, handling various $N$ ISOs and $M$ CFAs would require many $N \times M$ modes for the model. Our work takes the first step of handling different CFAs, and we hope our dataset enables future work on this problem.

\ifpeerreview \else
\section*{Acknowledgments}
The authors would like to thank Kosta Derpanis, Karmon Dhillon, Fereshteh Forghani, and Shirin Taghian, for their help with dataset materials and capture. 
\fi

\bibliographystyle{IEEEtran}
\bibliography{main}

\begin{thebibliography}{10}
\providecommand{\url}[1]{#1}
\csname url@samestyle\endcsname
\providecommand{\newblock}{\relax}
\providecommand{\bibinfo}[2]{#2}
\providecommand{\BIBentrySTDinterwordspacing}{\spaceskip=0pt\relax}
\providecommand{\BIBentryALTinterwordstretchfactor}{4}
\providecommand{\BIBentryALTinterwordspacing}{\spaceskip=\fontdimen2\font plus
\BIBentryALTinterwordstretchfactor\fontdimen3\font minus \fontdimen4\font\relax}
\providecommand{\BIBforeignlanguage}[2]{{%
\expandafter\ifx\csname l@#1\endcsname\relax
\typeout{** WARNING: IEEEtran.bst: No hyphenation pattern has been}%
\typeout{** loaded for the language `#1'. Using the pattern for}%
\typeout{** the default language instead.}%
\else
\language=\csname l@#1\endcsname
\fi
#2}}
\providecommand{\BIBdecl}{\relax}
\BIBdecl

\bibitem{isp1}
H.~C. Karaimer and M.~S. Brown, ``A software platform for manipulating the camera imaging pipeline,'' in \emph{ECCV}, 2016.

\bibitem{quadbayerexists}
A.~Ignatov, R.~Timofte, S.~Liu, C.~Feng, F.~Bai, X.~Wang, L.~Lei, Z.~Yi, Y.~Xiang, Z.~Liu \emph{et~al.}, ``Learned smartphone {ISP} on mobile {GPU}s with deep learning, mobile {AI} \& {AIM},'' in \emph{ECCV}, 2022.

\bibitem{nonabayerexists}
M.~Cho, H.~Lee, H.~Je, K.~Kim, D.~Ryu, and A.~No, ``Pynet-{Q}$\times$ {Q}: An efficient pynet variant for {Q}$\times$ {Q} {Bayer} pattern demosaicing in {CMOS} image sensors,'' \emph{IEEE Access}, 2023.

\bibitem{remosaic1}
J.~Jia, H.~Sun, X.~Liu, L.~Xiao, Q.~Xu, and G.~Zhai, ``Learning rich information for quad bayer remosaicing and denoising,'' in \emph{ECCV}, 2022.

\bibitem{remosaic2}
Y.~Kim, J.~Lee, S.~Kim, J.~Bang, D.~Hong, T.~Kim, and J.~Yim, ``Camera image quality tradeoff processing of image sensor re-mosaic using deep neural network,'' \emph{Electronic Imaging}, 2021.

\bibitem{bjdd}
S.~Sharif, R.~A. Naqvi, and M.~Biswas, ``Beyond joint demosaicking and denoising: An image processing pipeline for a pixel-bin image sensor,'' 2021.

\bibitem{sagan}
------, ``Sagan: Adversarial spatial-asymmetric attention for noisy nona-bayer reconstruction,'' in \emph{BMVC}, 2021.

\bibitem{jdndm}
W.~Xing and K.~Egiazarian, ``End-to-end learning for joint image demosaicing, denoising and super-resolution,'' in \emph{CVPR}, 2021.

\bibitem{zoom}
J.~W. Nash, A.~A. Golikeri, and N.~K.~S. Ravirala, ``Depth-based zoom function using multiple cameras,'' 2019, {US} Patent 10,389,948.

\bibitem{model-based-1}
H.~S. Malvar, L.-w. He, and R.~Cutler, ``High-quality linear interpolation for demosaicing of bayer-patterned color images,'' in \emph{ICASS}, vol.~3, 2004.

\bibitem{gc1}
K.~Hirakawa and T.~W. Parks, ``Adaptive homogeneity-directed demosaicing algorithm,'' \emph{TIP}, vol.~14, no.~3, 2005.

\bibitem{model-based-3}
L.~Zhang and X.~Wu, ``Color demosaicking via directional linear minimum mean square-error estimation,'' \emph{TIP}, vol.~14, no.~12, 2005.

\bibitem{cnns}
Z.~Li, F.~Liu, W.~Yang, S.~Peng, and J.~Zhou, ``A survey of convolutional neural networks: analysis, applications, and prospects,'' \emph{TNNLS}, 2021.

\bibitem{alexnet}
A.~Krizhevsky, I.~Sutskever, and G.~E. Hinton, ``Imagenet classification with deep convolutional neural networks,'' in \emph{NeurIPS}, 2012.

\bibitem{autoencoder}
P.~Baldi, ``Autoencoders, unsupervised learning, and deep architectures,'' in \emph{ICML Workshops}, 2012.

\bibitem{wang2014multilayer}
Y.-Q. Wang, ``A multilayer neural network for image demosaicking,'' in \emph{ICIP}, 2014.

\bibitem{tan2017color}
R.~Tan, K.~Zhang, W.~Zuo, and L.~Zhang, ``Color image demosaicking via deep residual learning,'' in \emph{ICME}, vol.~2, no.~4, 2017.

\bibitem{dpn}
I.~Kim, S.~Song, S.~Chang, S.~Lim, and K.~Guo, ``Deep image demosaicing for submicron image sensors,'' \emph{Electronic Imaging}, vol.~32, 2019.

\bibitem{sidd}
A.~Abdelhamed, S.~Lin, and M.~S. Brown, ``A high-quality denoising dataset for smartphone cameras,'' in \emph{CVPR}, 2018.

\bibitem{paris}
M.~Gharbi, G.~Chaurasia, S.~Paris, and F.~Durand, ``Deep joint demosaicking and denoising,'' \emph{ACM TOG}, vol.~35, no.~6, 2016.

\bibitem{kokkinos2018deep}
F.~Kokkinos and S.~Lefkimmiatis, ``Deep image demosaicking using a cascade of convolutional residual denoising networks,'' in \emph{ECCV}, 2018.

\bibitem{sgnet}
L.~Liu, X.~Jia, J.~Liu, and Q.~Tian, ``Joint demosaicing and denoising with self guidance,'' in \emph{CVPR}, 2020.

\bibitem{rcan}
Y.~Zhang, K.~Li, K.~Li, L.~Wang, B.~Zhong, and Y.~Fu, ``Image super-resolution using very deep residual channel attention networks,'' in \emph{ECCV}, 2018.

\bibitem{seeinginthedark}
C.~Chen, Q.~Chen, J.~Xu, and V.~Koltun, ``Learning to see in the dark,'' in \emph{CVPR}, 2018.

\bibitem{pixelshift200}
G.~Qian, Y.~Wang, J.~Gu, C.~Dong, W.~Heidrich, B.~Ghanem, and J.~S. Ren, ``Rethinking learning-based demosaicing, denoising, and super-resolution pipeline,'' in \emph{ICCP}, 2022.

\bibitem{remosaic_challenge}
Q.~Yang, G.~Yang, J.~Jiang, C.~Li, R.~Feng, S.~Zhou, W.~Sun, Q.~Zhu, C.~C. Loy, J.~Gu \emph{et~al.}, ``Mipi 2022 challenge on {Quad-Bayer} re-mosaic: Dataset and report,'' in \emph{ECCV}, 2022.

\bibitem{drunet}
K.~Zhang, Y.~Li, W.~Zuo, L.~Zhang, L.~Van~Gool, and R.~Timofte, ``Plug-and-play image restoration with deep denoiser prior,'' \emph{TPAMI}, vol.~44, no.~10, 2021.

\bibitem{unet}
O.~Ronneberger, P.~Fischer, and T.~Brox, ``U-net: Convolutional networks for biomedical image segmentation,'' in \emph{MICCAI}, 2015.

\bibitem{resnet}
K.~He, X.~Zhang, S.~Ren, and J.~Sun, ``Deep residual learning for image recognition,'' in \emph{CVPR}, 2016.

\bibitem{remosaic_doesnt_work}
S.~Gil, O.~Kim, E.~Yong, S.-S. Kim, and J.~Yim, ``Image distortion inference based on correlation between line pattern and character,'' \emph{Electronic Imaging}, vol.~34, no.~9, 2022.

\bibitem{KLAP}
H.~Lee, D.~Park, W.~Jeong, K.~Kim, H.~Je, D.~Ryu, and S.~Y. Chun, ``Efficient unified demosaicing for bayer and non-bayer patterned image sensors,'' in \emph{ICCV}, 2023.

\bibitem{gti}
{GTI}, ``{GTI Graphics Technology Inc.}'' \url{https://www.gtilite.com/products/color-matching-systems/gti-minimatcher-series/}, 2024, accessed: 2024-10-15.

\bibitem{colom2013analysis}
M.~Colom and A.~Buades, ``Analysis and extension of the ponomarenko et al. method, estimating a noise curve from a single image,'' \emph{Image Processing On Line}, vol.~3, pp. 173--197, 2013.

\bibitem{plotz}
T.~Plotz and S.~Roth, ``Benchmarking denoising algorithms with real photographs,'' in \emph{CVPR}, 2017.

\bibitem{brooks2019unprocessing}
T.~Brooks, B.~Mildenhall, T.~Xue, J.~Chen, D.~Sharlet, and J.~T. Barron, ``Unprocessing images for learned raw denoising,'' in \emph{CVPR}, 2019.

\bibitem{nafnet}
L.~Chen, X.~Chu, X.~Zhang, and J.~Sun, ``Simple baselines for image restoration,'' in \emph{ECCV}.\hskip 1em plus 0.5em minus 0.4em\relax Springer, 2022.

\bibitem{bradski2000opencv}
G.~Bradski, A.~Kaehler \emph{et~al.}, ``Open{CV},'' \emph{Dr. Dobb's Journal of Software Tools}, vol.~3, no.~2, 2000.

\bibitem{menon2006demosaicing}
D.~Menon, S.~Andriani, and G.~Calvagno, ``Demosaicing with directional filtering and a posteriori decision,'' \emph{IEEE TIP}, 2006.

\bibitem{deadpixel1}
C.-W. Chen, C.-Y. Cho, Y.-F. Sun, T.-M. Chen, and C.-L. Su, ``Low complexity photo sensor dead pixel detection algorithm,'' in \emph{APCCS}, 2012.

\bibitem{mosleh2024non}
A.~Mosleh, L.~Zhao, A.~Singh, J.~Han, A.~Punnappurath, M.~A. Brubaker, J.~Choe, and M.~S. Brown, ``Non-parametric sensor noise modeling and synthesis,'' in \emph{ECCV}.\hskip 1em plus 0.5em minus 0.4em\relax Springer, 2024.

\end{thebibliography}

\ifpeerreview \else


\begin{IEEEbiography}[{\includegraphics[width=1in,height=1.25in,clip,keepaspectratio]{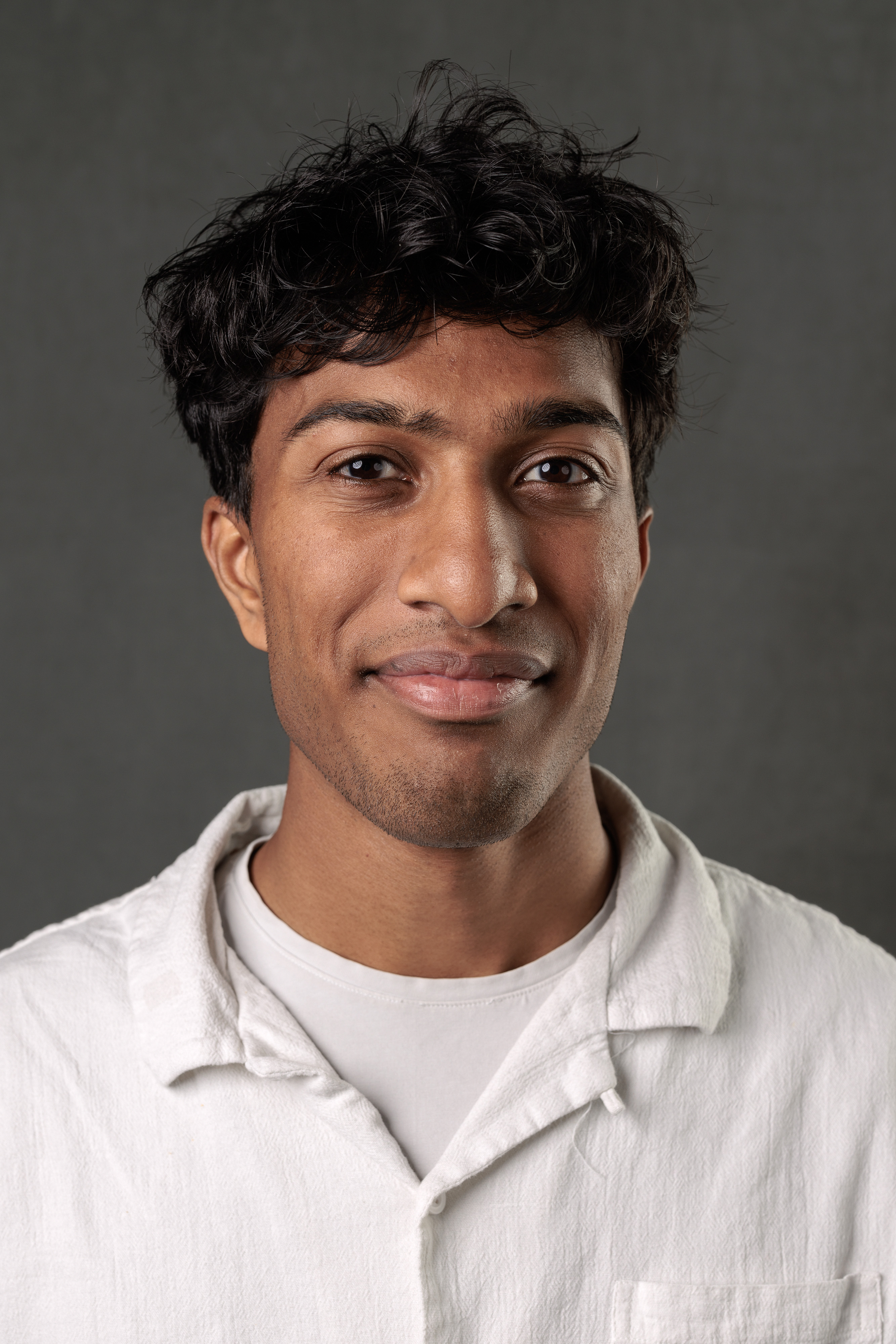}}]{SaiKiran Tedla}
SaiKiran (Sai) Tedla is a PhD student in the Department of Electrical Engineering and Computer Science at York University, advised by Michael S. Brown. Currently, he is an intern at Samsung AI Center Toronto and has previously interned at Adobe. His research is focused on computational photography and camera pipeline design. 
\end{IEEEbiography}

\begin{IEEEbiography}[{\includegraphics[width=1in,height=1.25in,clip,keepaspectratio]{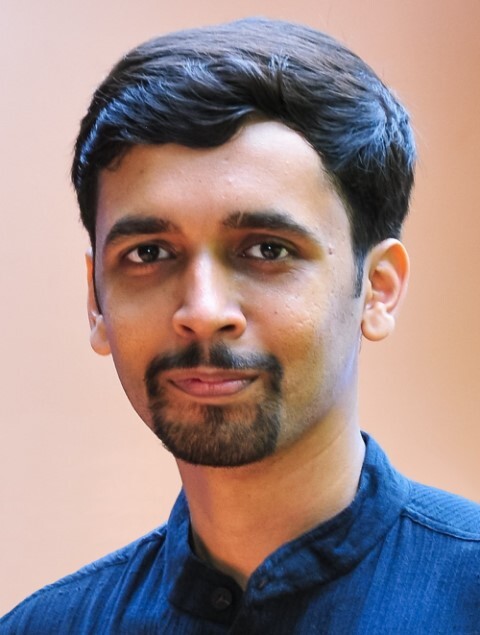}}]{Abhijith Punnappurath}
Abhijith Punnappurath is a research scientist at AI Center-Toronto, Samsung Electronics. He was a Postdoctoral Fellow at the Electrical Engineering and Computer Science department, York University, Toronto, Canada. He received his Ph.D. degree from the Electrical Engineering department, Indian Institute of Technology Madras, India, in 2017. His research interests lie in the areas of low-level computer vision, computational photography, and machine learning.
\end{IEEEbiography}

\begin{IEEEbiography}[{\includegraphics[width=1in,height=1in,clip,keepaspectratio]{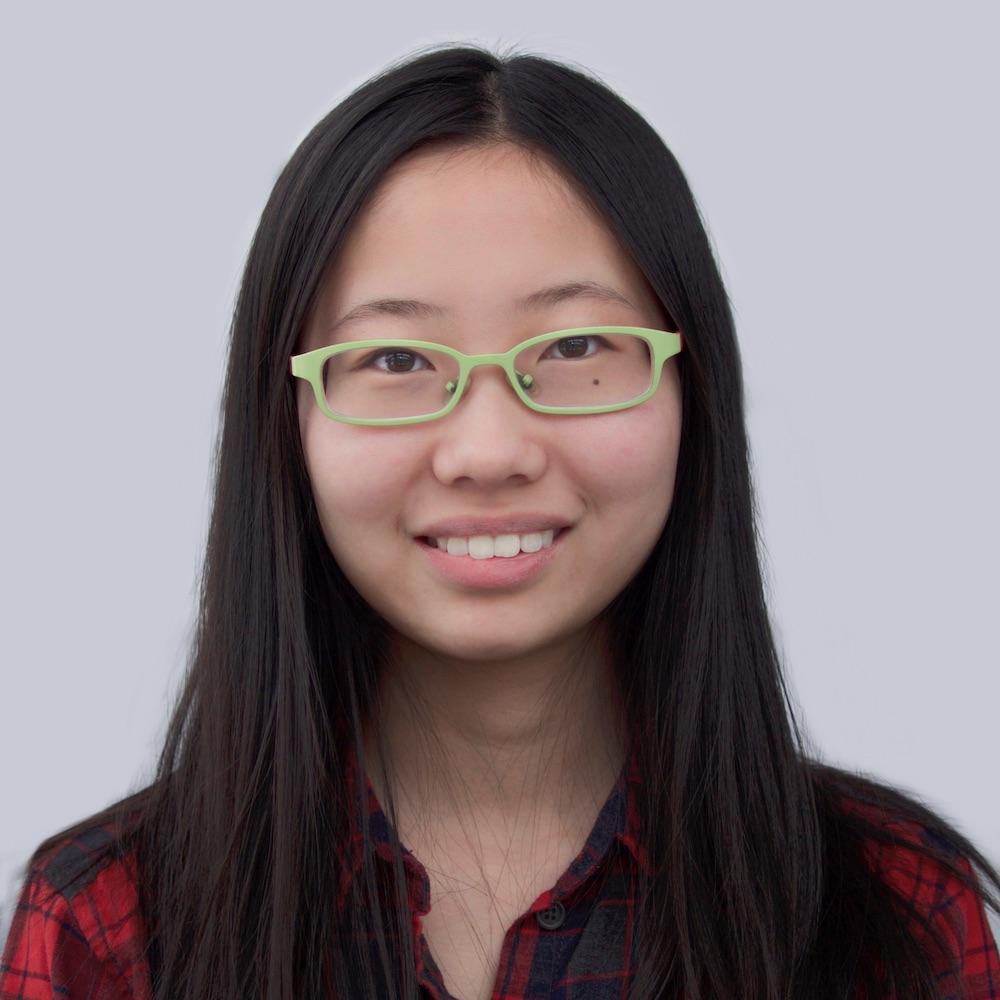}}]{Luxi Zhao}
Luxi Zhao is a research engineer at AI Center-Toronto, Samsung Electronics. She received the Bachelor of Applied Science degree in computer engineering from The University of British Columbia, in 2020, and the Master of Science degree in applied computing from the University of Toronto, in 2022. Her research interests include low-level computer vision, computational photography, and machine learning.
\end{IEEEbiography}

\begin{IEEEbiography}[{\includegraphics[width=1in,height=1in,clip,keepaspectratio]{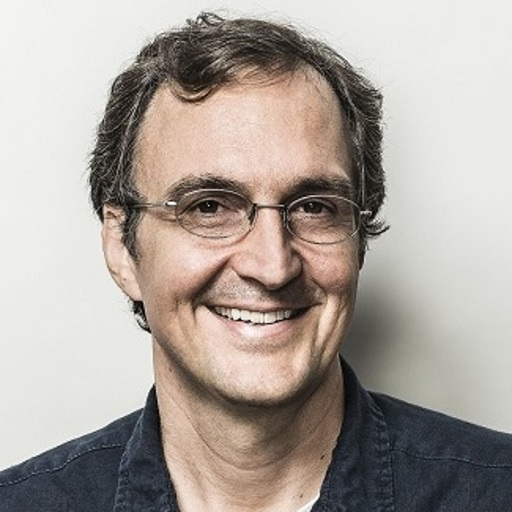}}]{Michael S. Brown}
Michael S. Brown is the lab director at the AI Center-Toronto, Samsung Electronics. Dr. Brown is also a professor and Canada Research Chair at York University. His research is focused on in-camera color processing and data synthesis for low-level computer vision tasks. Dr. Brown is a fellow of the IEEE.
\end{IEEEbiography}




\fi

\newpage 

\maketitlesupplementary
\setcounter{section}{0}

\section{Hard demosaicing dataset}
\label{sec:supp_dataset}
\begin{figure*}[]
  \centering
  \includegraphics[width=1.0\linewidth]{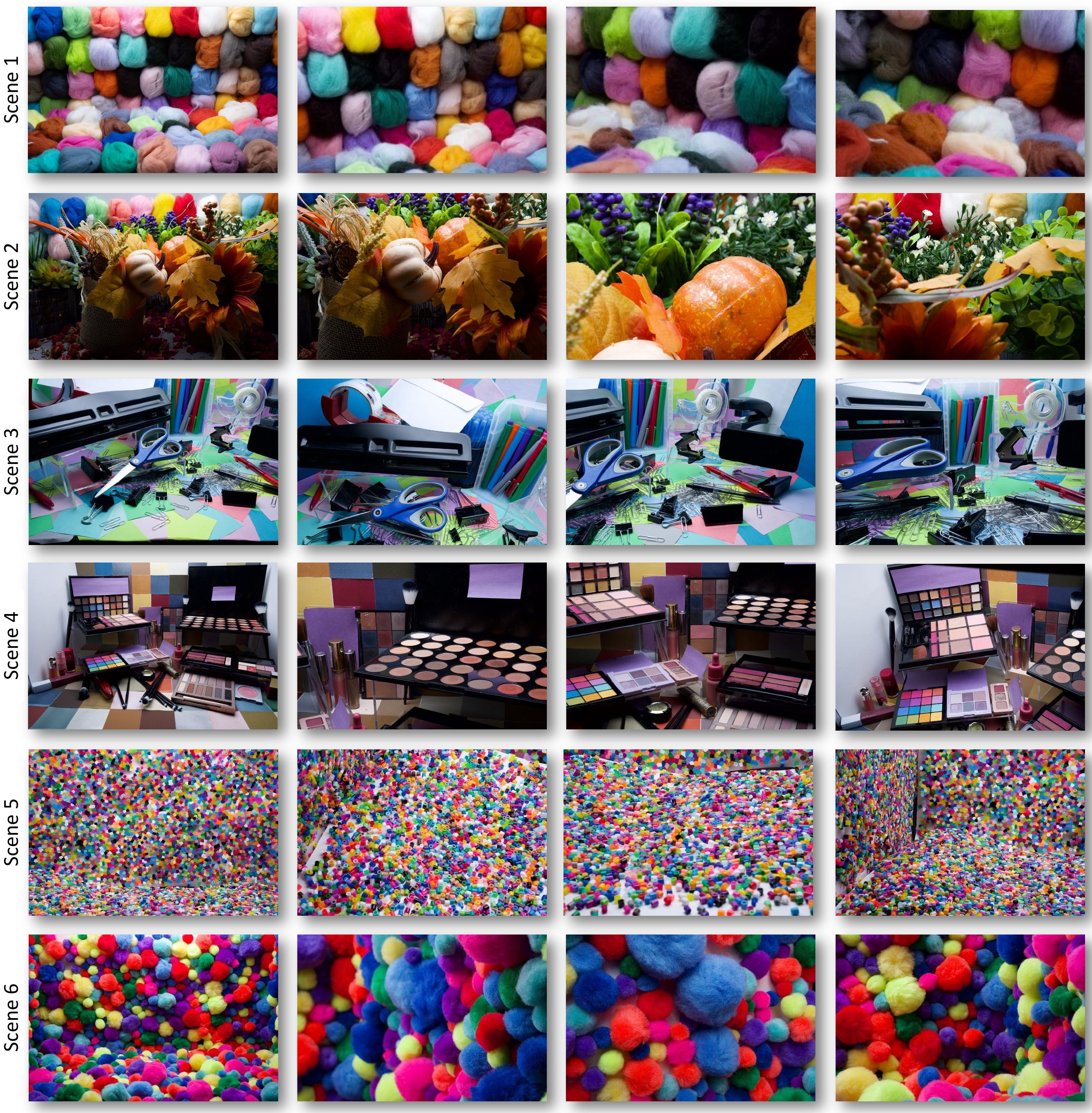}
  \caption[Scenes]{Scenes 1-6 of our dataset with four different views. }\label{fig:supp_dataset_1}
\end{figure*}

\begin{figure*}[]
  \centering
  \includegraphics[width=1.0\linewidth]{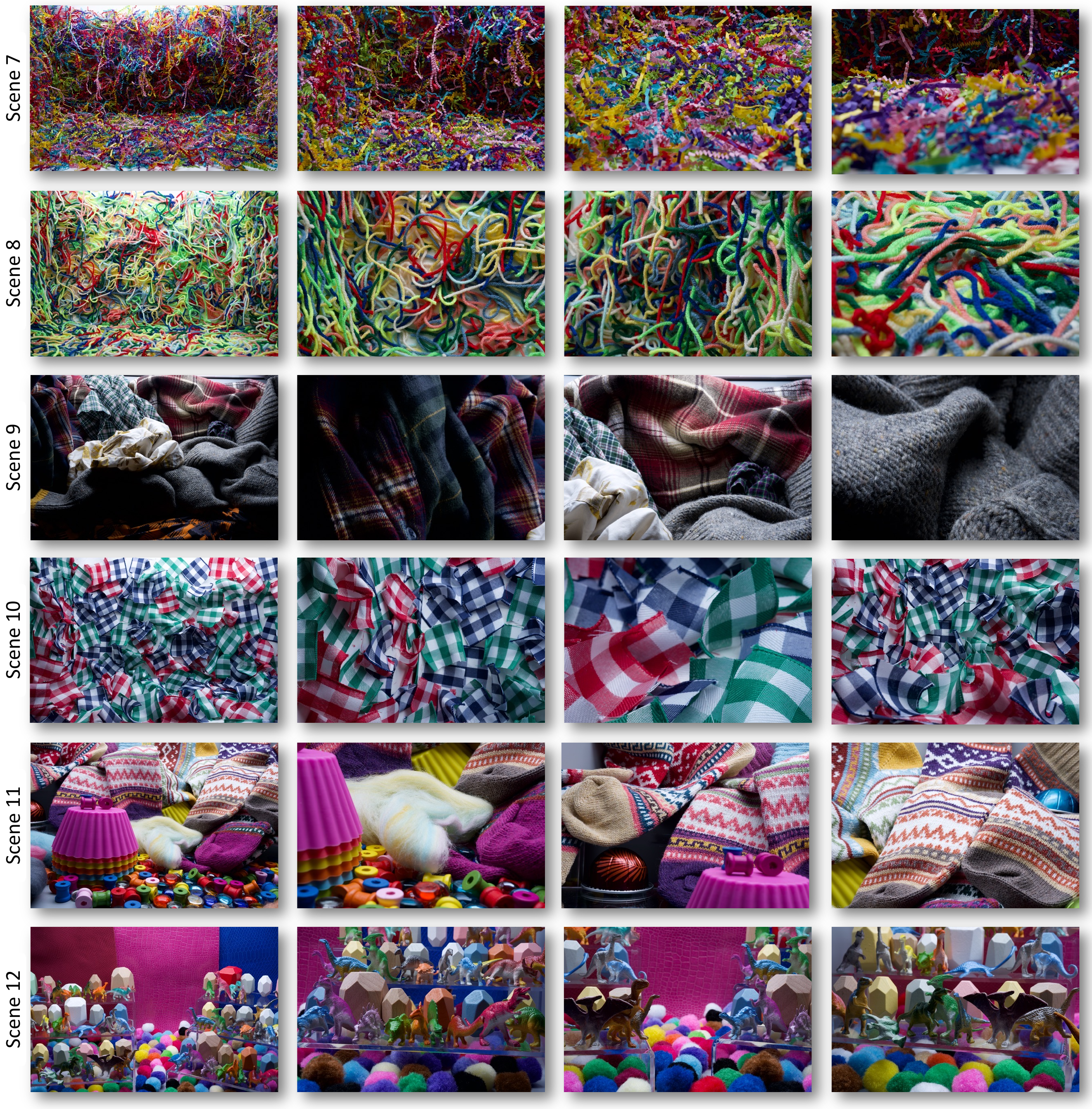}
  \caption[Scenes]{Scenes 7-12 of our dataset with four different views. }\label{fig:supp_dataset_2}
\end{figure*}

\begin{figure*}[]
  \centering
  \includegraphics[width=1.0\linewidth]{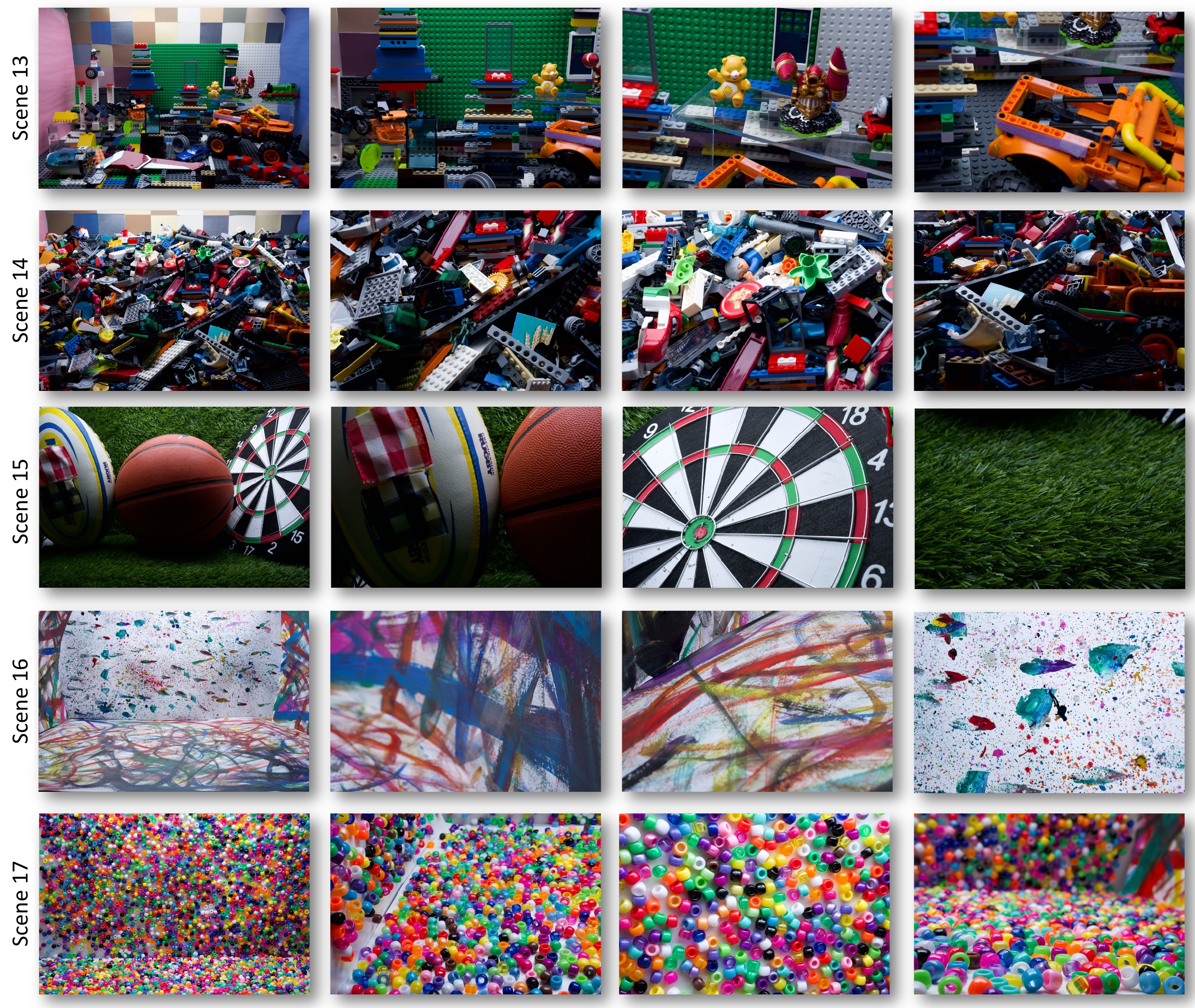}
  \caption[Scenes]{Scenes 13-17 of our dataset with four different views.}\label{fig:supp_dataset_3}
\end{figure*}

Our hard demosaicing dataset (HDD) was captured carefully to contain scenes consisting of high-frequency detail. This was done because thin edges and small details are challenging for demosaicing algorithms.  To achieve our goal, we used various materials and textures. For each scene, we also captured varying numbers of views, as some scenes had more interesting content than others. We visualize four views of each scene in Figures~\ref{fig:supp_dataset_1}-\ref{fig:supp_dataset_3}.

We also conducted experiments to show the value of using RAW images for training  (instead of sRGB images) and the use of hard patch mining. Specifically, we compare JDNDM~\cite{jdndm} trained on our RAW images versus the original model trained on sRGB. Since the pre-trained JDMDM used Gaussian noise, we use a grid search to find the optimal noise level parameter per ISO level in our dataset that matched the Gaussian noise used in the original trained JDNDM. Additionally, we report results when training with all patches and ``easy'' patches (patches that were not labeled as hard). For these comparisons, we make sure that all compared models are trained for at least the same number of iterations as the model trained with hard patches. We report the results in Table~\ref{tab:supp_dataset}. Additionally, we report results for a similar experiment with our unified model, ESUM, in Table~\ref{tab:supp_esum_dataset}.

We noticed that the pre-trained model with sRGB data did not perform well on our RAW images. Additionally, training with hard patches was better than training with all patches at each of the four ISO levels and was most noticeable at ISO 400. At the other ISO levels, results were only slightly better for hard patch training and this might be because hard patches are a subset of all patches. However, we see a significant gap between training with hard patches and training with easy patches. This illustrates that hard patches in our dataset are valuable for training a model that demosaics RAW images with high texture.

\begin{figure}[h]
  \centering
  \includegraphics[width=1.0\linewidth]{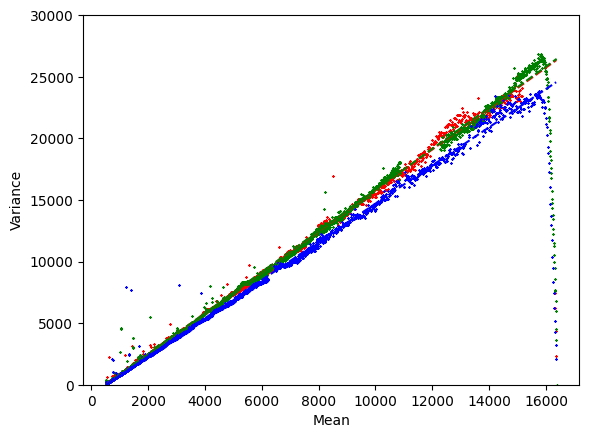}
  \caption[NoiseModel400]{We visualize our noise model calibration for ISO 400. At each intensity value, we perform calibration to find the corresponding variance. The Poisson-Gaussian model is given by the lines fitted through these points.}\label{fig:noise_model_400}
\end{figure}

\begin{figure}[h]
  \centering
  \includegraphics[width=1.0\linewidth]{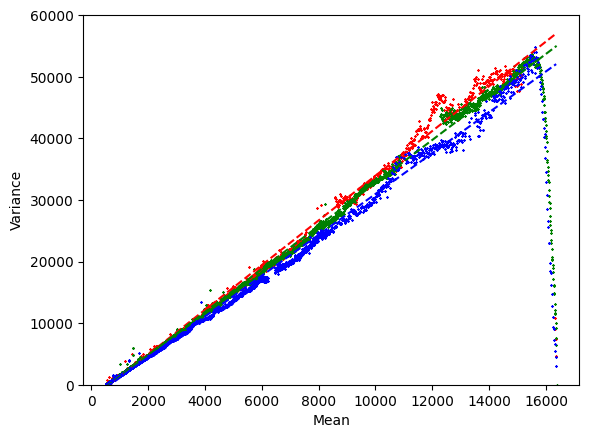}
  \caption[NoiseModel800]{We visualize our noise model calibration for ISO 800. At each intensity value, we perform calibration to find the corresponding variance. The Poisson-Gaussian model is given by the lines fitted through these points.}\label{fig:noise_model_800}
\end{figure}

\begin{figure}[h]
  \centering
  \includegraphics[width=1.0\linewidth]{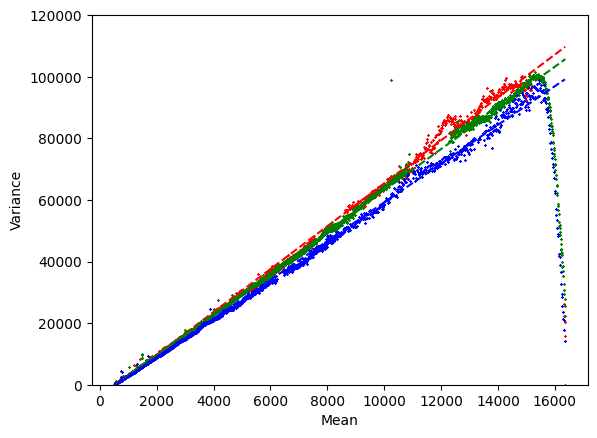}
  \caption[NoiseModel1600]{We visualize our noise model calibration for ISO 1600. At each intensity value, we perform calibration to find the corresponding variance. The Poisson-Gaussian model is given by the lines fitted through these points.}\label{fig:noise_model_1600}
\end{figure}

\begin{figure}[h]
  \centering
  \includegraphics[width=1.0\linewidth]{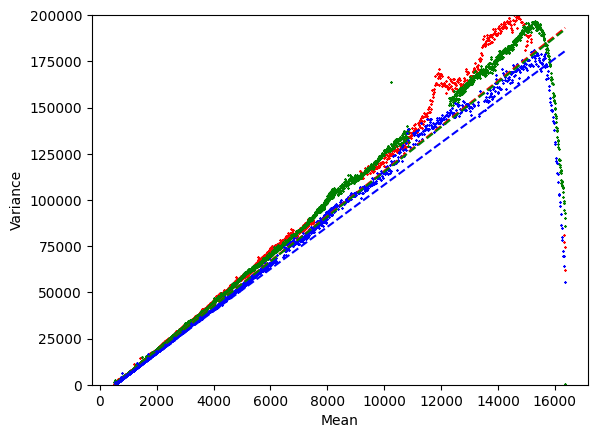}
  \caption[NoiseModel3200]{We visualize our noise model calibration for ISO 3200. At each intensity value, we perform calibration to find the corresponding variance. The Poisson-Gaussian model is given by the lines fitted through these points.}\label{fig:noise_model_3200}
\end{figure}

Finally, we calibrated heteroschedastic Poisson-Gaussian noise by following the procedure in Plotz et al~\cite{plotz}. We additionally calibrated non-parametric models~\cite{mosleh2024non}. Both noise models are similar and are calibrated by finding the variance of noise at each intensity value from a set of images (we use 30 images at 6 EVs). In the parametric (Poisson-Gaussian) model, a line is fit through these calibrated points and then variances are sampled according to this fitting. In the non-parametric model, the variances are directly sampled from the calibration. Although, we did not use these non-parametric models in our experiments, we include them with our dataset. Viusalization of the parametric (Poisson-Gaussian) and non-parametric noise models at ISO 400 through ISO 3200 are shown in Figure~\ref{fig:noise_model_400}-\ref{fig:noise_model_3200}. 

\section{Unified models}
\label{sec:supp_unified_models}

\begin{figure*}[]
  \centering
  \includegraphics[width=0.99\linewidth]{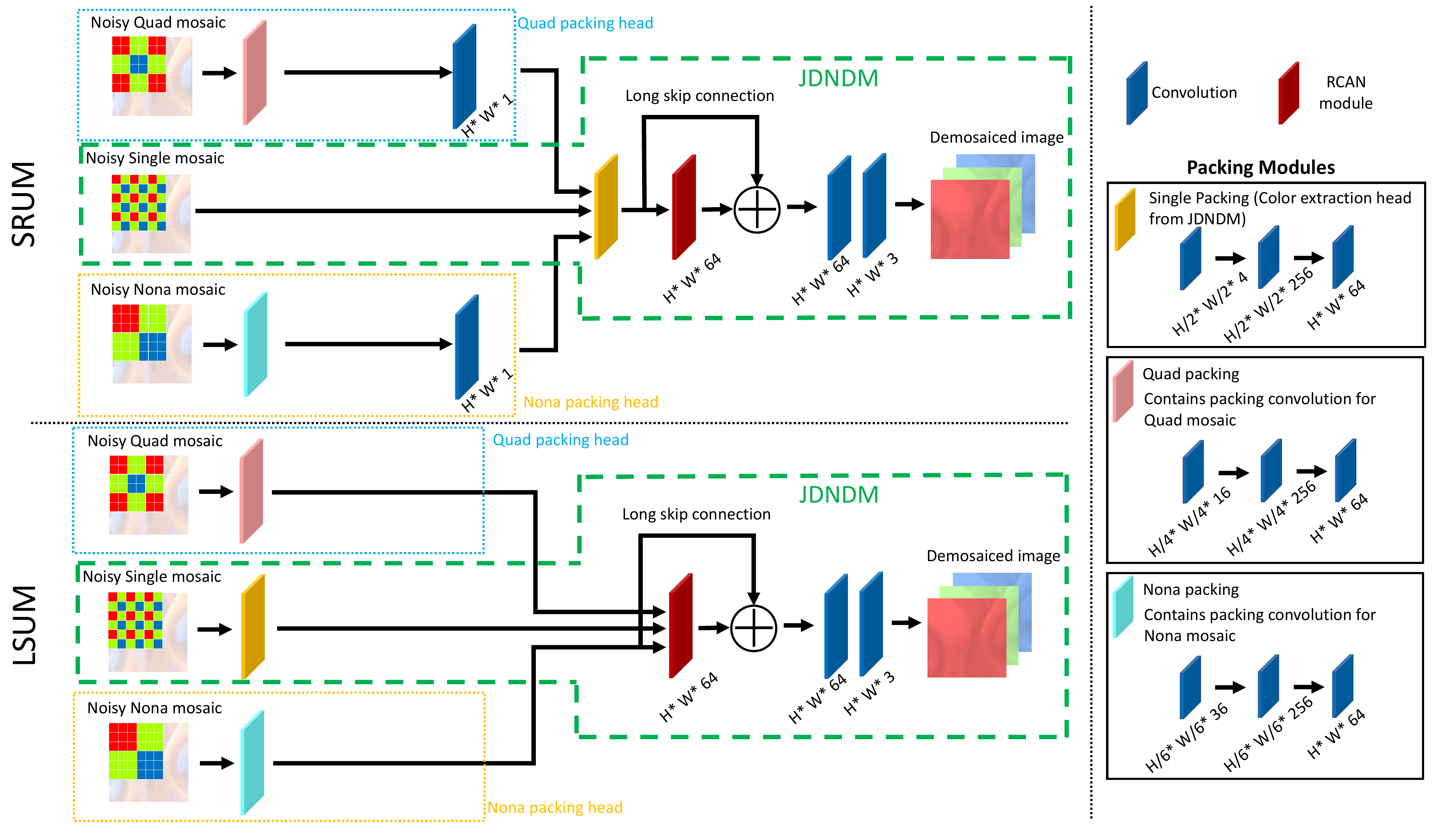}
  \caption[Model]{We illustrate two of the unified model approaches (SRUM and LSUM) with their packing head versions. Each packing head contains a ``packing'' convolution that packs the mosaic based on its pattern type.}\label{fig:supp_model}
\end{figure*}

In the main paper, we looked at SRUM and LSUM: two approaches for a unified joint demosaicing and denoising model. Both approaches use ``heads'' to deal with the different pattern types. We tried two approaches for the heads: shuffle and packing. We illustrated the shuffle head versions in the main paper. The packing head versions of SRUM and LSUM are illustrated in Figure~\ref{fig:supp_model}. The packing head contains a ``packing'' convolution that packs the mosaic based on its pattern type. Normally a ``packed mosaic'' is in the context of the Single-Bayer pattern and is size $H/2 \times W/2 \times 4$. However, for the new Quad-Bayer and Nona-Bayer patterns, the sizes are $H/4 \times W/4 \times 16$ and $H/6 \times W/6 \times 36$, respectively.

\section{Quantitative results}

In the main paper, we only report results on the hardest 25\% of patches, so we additionally include results on the full images in Tables~\ref{tab:full_individual_models} and ~\ref{tab:full_unified_models}. Additionally, these tables contain SSIM and Delta E to measure demosaicing performance. These results follow similar trends to the observations described in Section 4.2 and 5.2 of the main paper.

\section{Qualitative results}
\label{sec:supp_qualitative_results}

\begin{figure*}[h]
  \centering
  \includegraphics[width=0.89\linewidth]{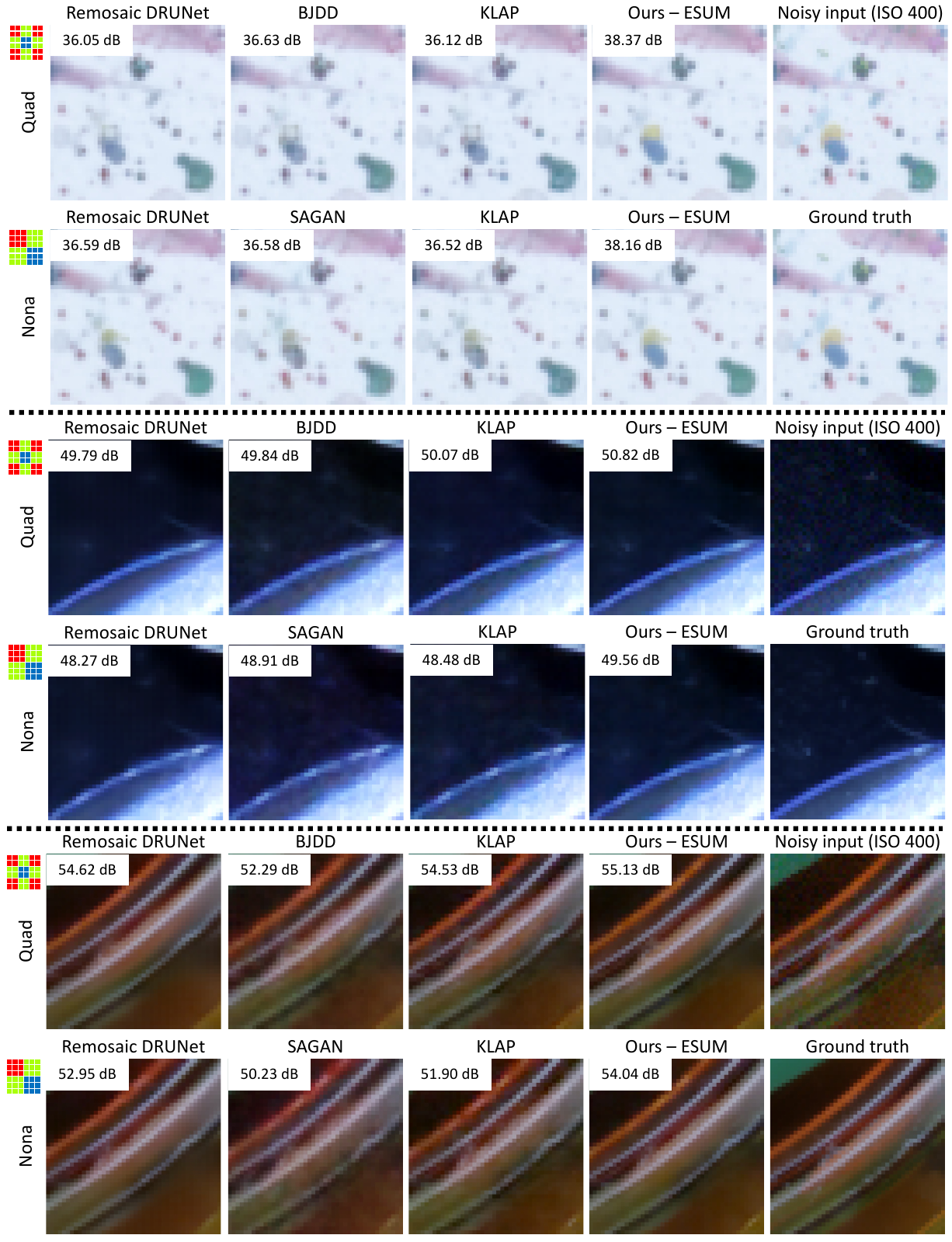}
  \caption{Qualitative results comparing our embedding-based unified model, ESUM, with existing individual demosaicing methods (remosaic with DRUNet~\cite{drunet}, BJDD~\cite{bjdd}, SAGAN~\cite{sagan}) and a unified method, KLAP~\cite{KLAP}, for Quad-Bayer and Nona-Bayer mosaics. We show three patches at ISO 400 (noisy input is before mosaic sampling). PSNR is reported in RAW, but visualized images are rendered by an ISP~\cite{sidd}.} \label{fig:supp_qualitative_results_ISO400}
\end{figure*}

\begin{figure*}[h]
  \centering
  \includegraphics[width=0.89\linewidth]{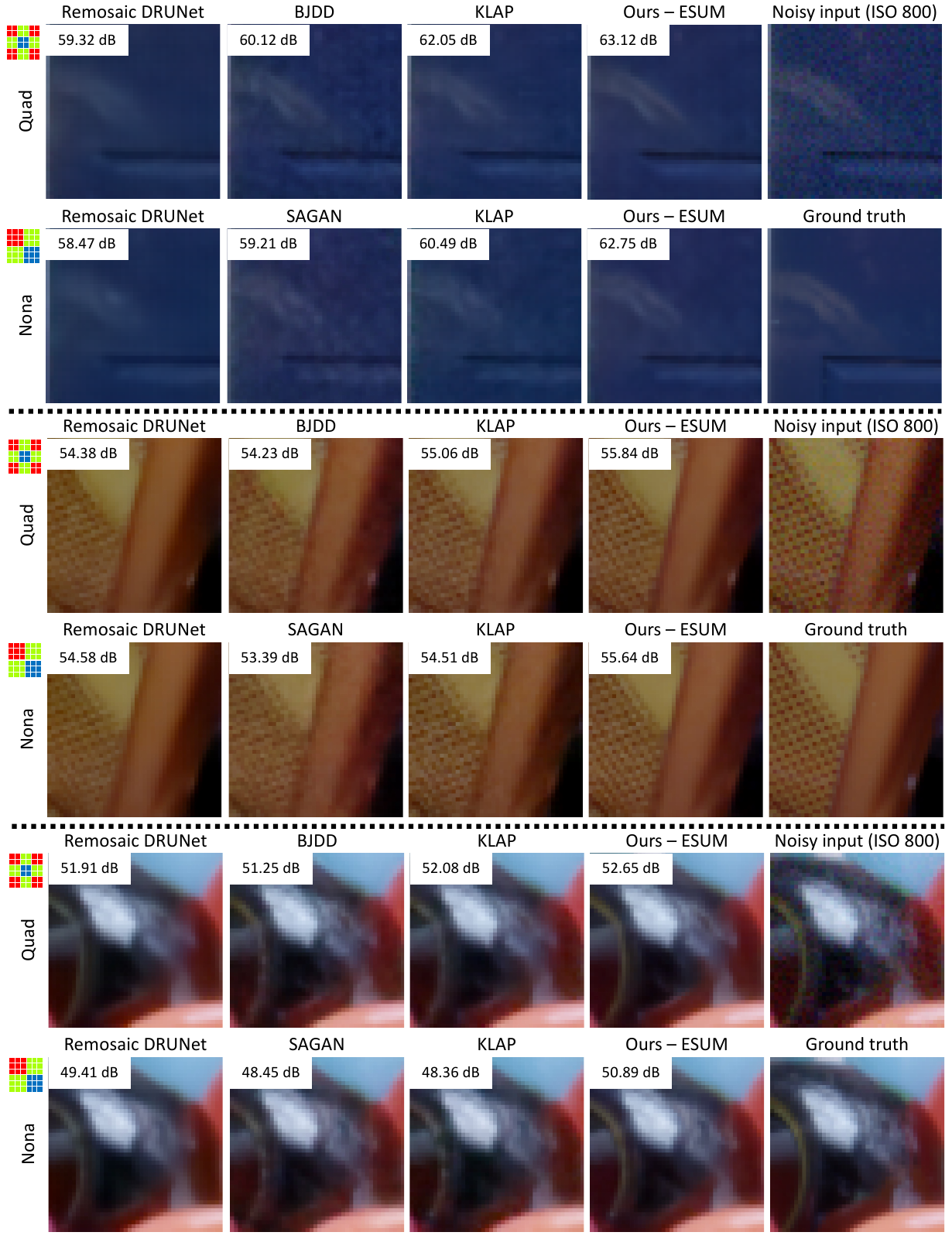}
  \caption{Qualitative results comparing our embedding-based unified model, ESUM, with existing individual demosaicing methods (remosaic with DRUNet~\cite{drunet}, BJDD~\cite{bjdd}, SAGAN~\cite{sagan}) and a unified method, KLAP~\cite{KLAP}, for Quad-Bayer and Nona-Bayer mosaics. We show three patches at ISO 800 (noisy input is before mosaic sampling). PSNR is reported in RAW, but visualized images are rendered by an ISP~\cite{sidd}.} \label{fig:supp_qualitative_results_ISO800}
\end{figure*}

\begin{figure*}[h]
  \centering
  \includegraphics[width=0.89\linewidth]{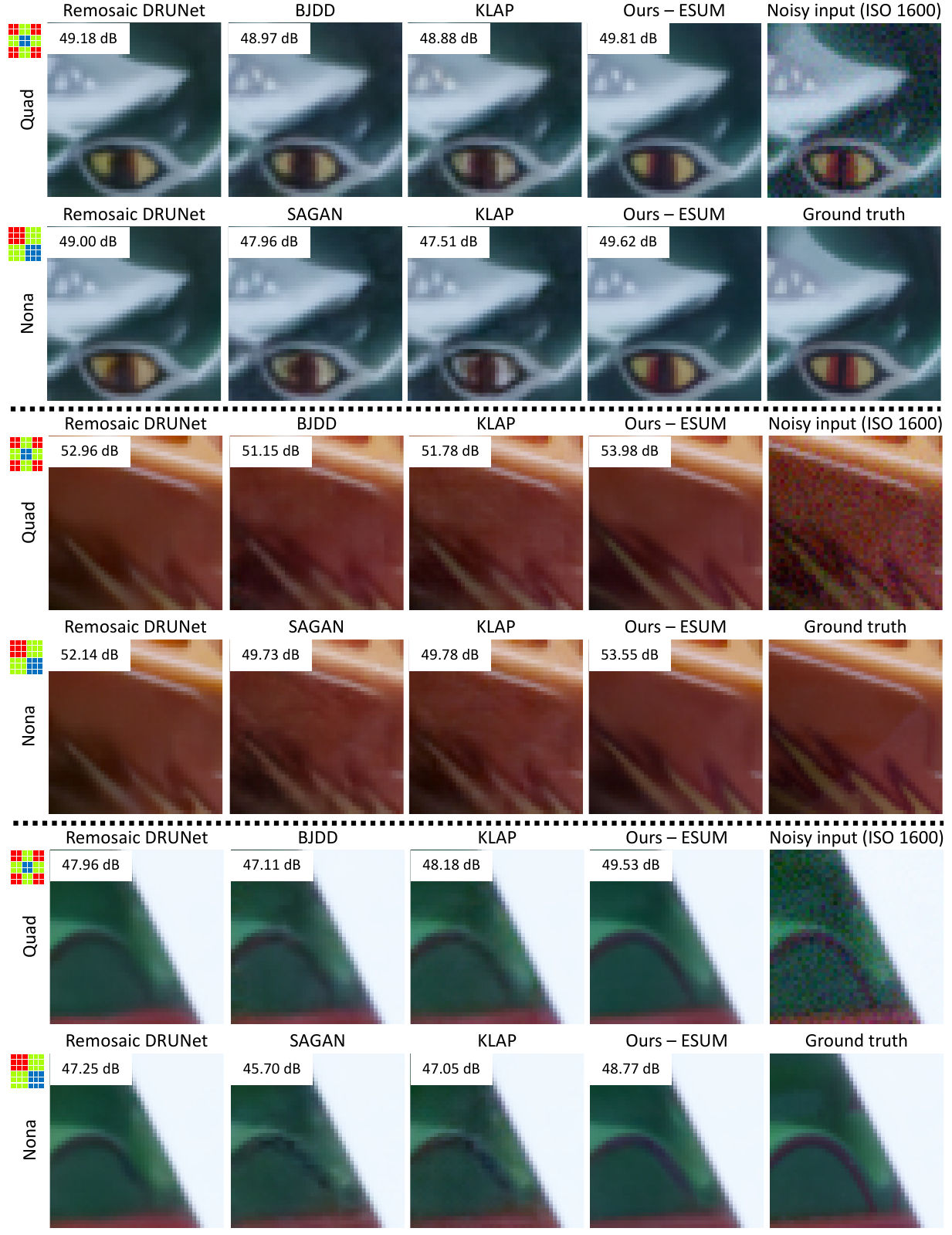}
  \caption{Qualitative results comparing our embedding-based unified model, ESUM, with existing individual demosaicing methods (remosaic with DRUNet~\cite{drunet}, BJDD~\cite{bjdd}, SAGAN~\cite{sagan}) and a unified method, KLAP~\cite{KLAP}, for Quad-Bayer and Nona-Bayer mosaics. We show three patches at ISO 1600 (noisy input is before mosaic sampling). PSNR is reported in RAW, but visualized images are rendered by an ISP~\cite{sidd}.} \label{fig:supp_qualitative_results_ISO1600}
\end{figure*}

\begin{figure*}[h]
  \centering
  \includegraphics[width=0.89\linewidth]{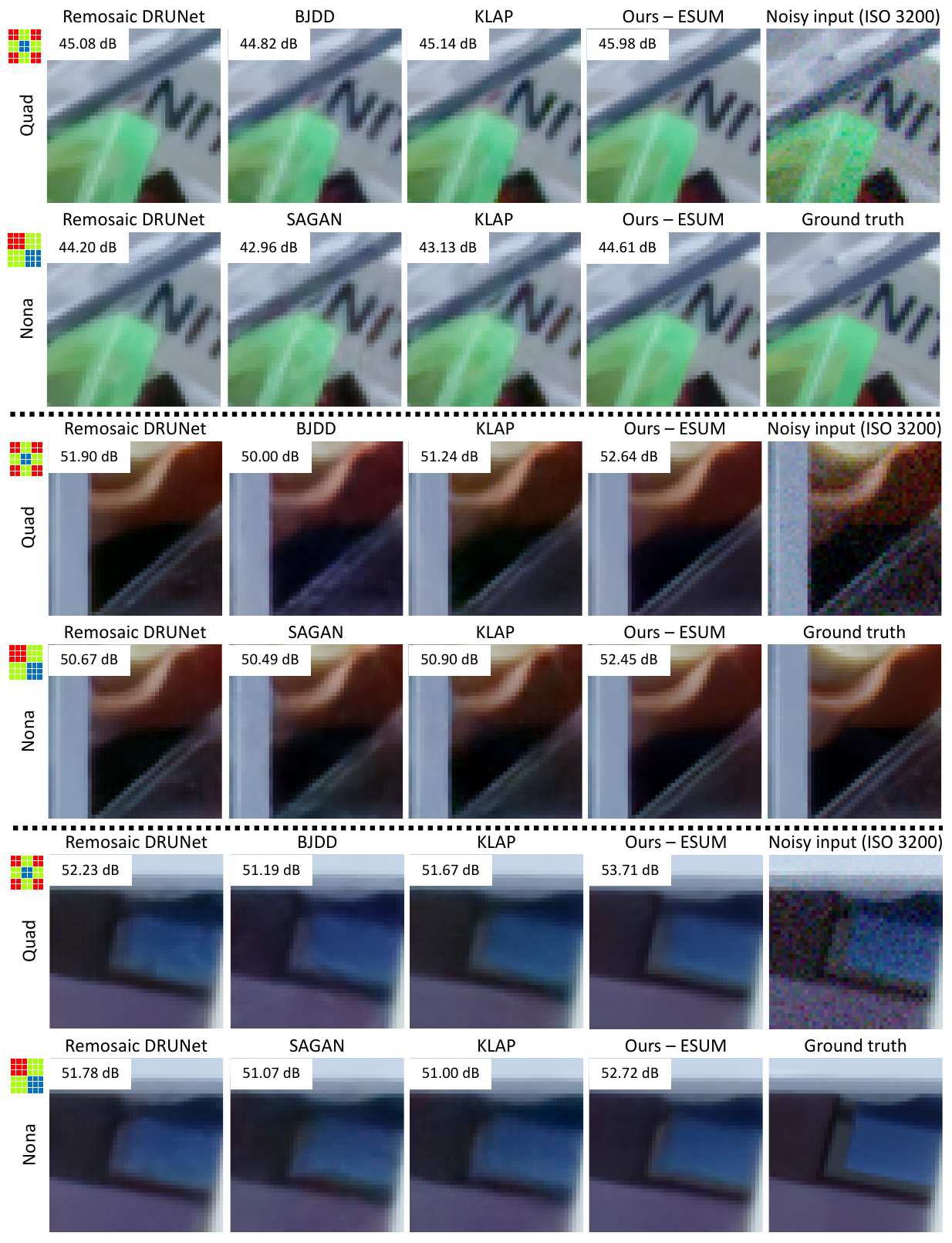}
  \caption{Qualitative results comparing our embedding-based unified model, ESUM, with existing individual demosaicing methods (remosaic with DRUNet~\cite{drunet}, BJDD~\cite{bjdd}, SAGAN~\cite{sagan}) and a unified method, KLAP~\cite{KLAP}, for Quad-Bayer and Nona-Bayer mosaics. We show three patches at ISO 3200 (noisy input is before mosaic sampling). PSNR is reported in RAW, but visualized images are rendered by an ISP~\cite{sidd}.} \label{fig:supp_qualitative_results_ISO3200}
\end{figure*}

\begin{figure*}[h]
  \centering
  \includegraphics[width=0.89\linewidth]{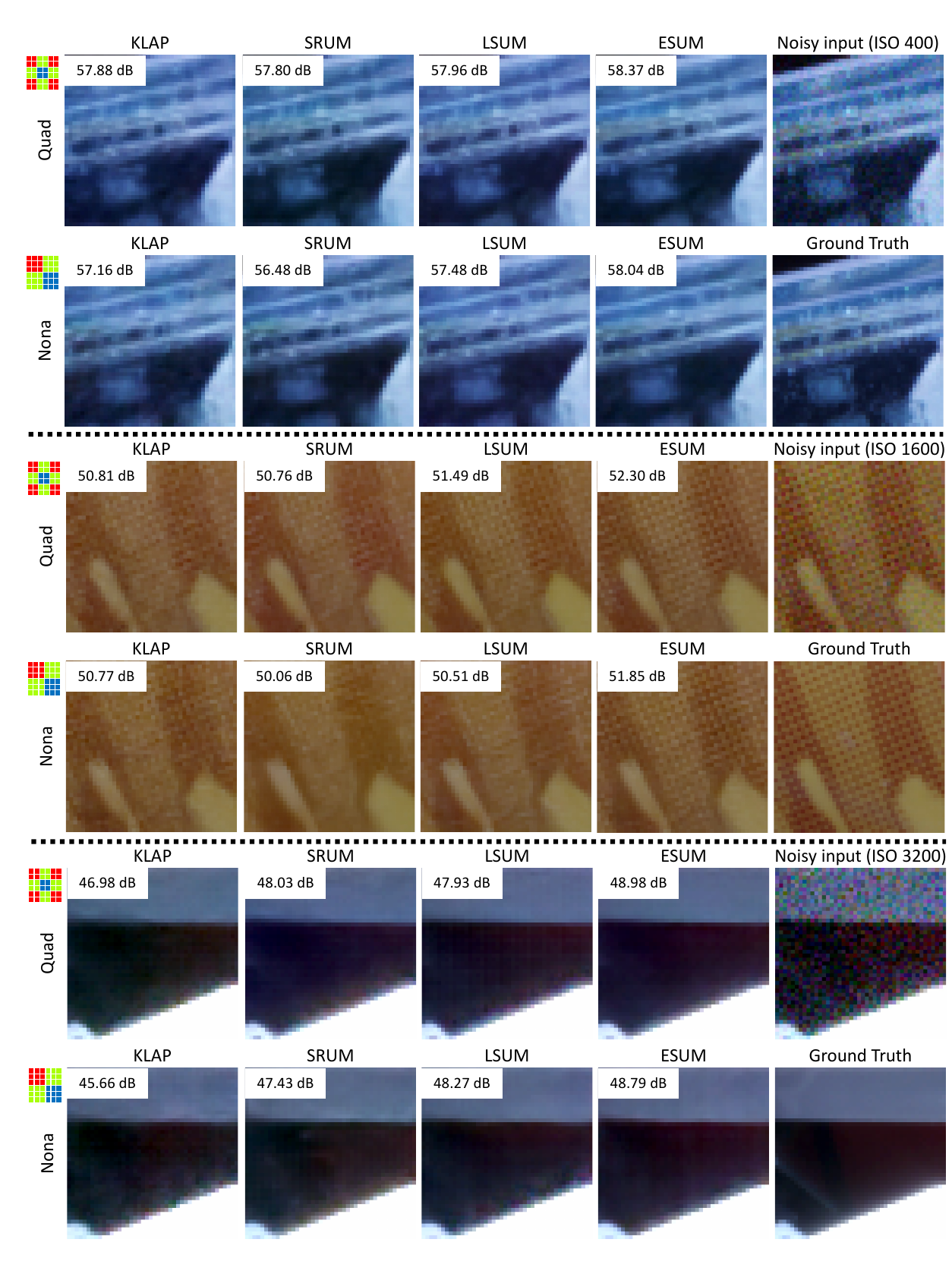}
  \caption{Qualitative results comparing our unified model approaches (SRUM, LSUM, ESUM) and another unified method, KLAP~\cite{KLAP}, for Quad-Bayer and Nona-Bayer mosaics. We show one patch at ISO 400, 1600, and 3200 (noisy input is before mosaic sampling). PSNR is reported in RAW, but visualized images are rendered by an ISP~\cite{sidd}.} \label{fig:supp_qualitative_results_unified}
\end{figure*}

Additional qualitative results of our model are provided in Figures~\ref{fig:supp_qualitative_results_ISO400}-\ref{fig:supp_qualitative_results_ISO3200}. We visualize three example patches at all four ISO levels tested. Additionally, Figure~\ref{fig:supp_qualitative_results_unified} contains qualitative results comparing the three unified model approaches.

\section{Additional implementation details}
For all experiments, we use manually defined random seeds. Additionally, all experiments are repeated three times and the average is reported in our tables.

\begin{table*}[t]
    \small
    \centering
    \caption[hj]{Results of Single-Bayer demosaicing with JDNDM on different datasets. We compare our model trained with hard patches from HDD with the pre-trained sRGB model from the authors. Additionally, we compare against training our model with all patches and the ``easy'' (not hard) patches in our dataset. We report on the test split of the hard patches and full images in our dataset at all 4 ISO levels. Green, yellow, and blue highlighting represents first, second, and third best, respectively.
    }
    \resizebox{0.7\textwidth}{!}{
        \renewcommand{\multirowsetup}{\centering}
        \setlength{\tabcolsep}{10pt}
        \begin{tabular}{l>{\centering\arraybackslash}cccc}
            \toprule \multicolumn{1}{l}{}                 & \multicolumn{4}{|c}{\textbf{Hard patches (PSNR) $\uparrow$ }}                                                                                                                          \\
            \cline{2-5}
            \multicolumn{1}{l}{\textbf{Training Dataset}} & \multicolumn{1}{|c|}{\textbf{ISO 400}}                        & \multicolumn{1}{c|}{\textbf{ISO 800}} & \multicolumn{1}{c|}{\textbf{ISO 1600}} & \multicolumn{1}{c}{\textbf{ISO 3200}} \\
            \toprule
            HDD - Hard Patches                            & \cellcolor{green!30}53.69                                     & \cellcolor{green!30}52.34             & \cellcolor{green!30}50.43              & \cellcolor{green!30}49.00             \\
            HDD - All Patches                             & \cellcolor{yellow!35}53.06                                    & \cellcolor{yellow!35}52.25            & \cellcolor{yellow!35}50.25             & \cellcolor{yellow!35}48.80            \\
            HDD - Easy Patches                            & \cellcolor{blue!30}51.71                                       & \cellcolor{blue!30}51.67               & \cellcolor{blue!30}49.48                & \cellcolor{blue!30}48.17               \\
            sRGB (Pre-trained JDNDM)                      & 49.14                                                         & 48.42                                 & 47.38                                  & 46.03                                 \\
            \toprule \multicolumn{1}{c}{}                 & \multicolumn{4}{c}{\textbf{Full images  (PSNR) $\uparrow$ }}                                                                                                                           \\
            \toprule
            HDD - Hard Patches                            & \cellcolor{green!30}56.36                                     & \cellcolor{green!30}55.01             & \cellcolor{green!30}53.04              & \cellcolor{green!30}51.74             \\
            HDD - All Patches                             & \cellcolor{yellow!35}55.77                                    & \cellcolor{yellow!35}54.94            & \cellcolor{yellow!35}53.02             & \cellcolor{yellow!35}51.54            \\
            HDD - Easy Patches                            & \cellcolor{blue!30}53.78                                       & \cellcolor{blue!30}53.82               & \cellcolor{blue!30}51.99                & \cellcolor{blue!30}50.66               \\
            sRGB (Pre-trained JDNDM)                      & 51.36                                                         & 50.67                                 & 49.67                                  & 48.37                                 \\
        \end{tabular}
    }

    \label{tab:supp_dataset}
\end{table*}

\begin{table*}[t]
    \small
    \centering
    \caption[hj]{Results of demosaicing with ESUM on different splits of our HDD dataset at ISO 400. We compare our model trained with hard, all, and ``easy'' (not hard) patches from HDD at. We report on the test split of the hard patches and full images. Green, yellow, and blue highlighting represents first, second, and third best, respectively.
    }
    \resizebox{0.5\textwidth}{!}{
        \renewcommand{\multirowsetup}{\centering}
        \setlength{\tabcolsep}{10pt}
        \begin{tabular}{l>{\centering\arraybackslash}ccc}
            \toprule \multicolumn{1}{l}{}                 & \multicolumn{3}{|c}{\textbf{Hard patches (PSNR) $\uparrow$ }}                                                                                                                          \\
            \cline{2-4}
            \multicolumn{1}{l}{\textbf{Training Dataset}} & \multicolumn{1}{|c|}{Single}                        & \multicolumn{1}{c|}{Quad} & \multicolumn{1}{c|}{Nona}\\
            \toprule
            HDD - Hard Patches                            & \cellcolor{green!30}53.75                                    & \cellcolor{green!30}52.68             & \cellcolor{green!30}51.96      \\
            HDD - All Patches                             & \cellcolor{yellow!35}53.13                                   & \cellcolor{yellow!35}51.99            & \cellcolor{yellow!35}51.15      \\
            HDD - Easy Patches                            & \cellcolor{blue!30}52.48                                       & \cellcolor{blue!30}51.39               & \cellcolor{blue!30}50.55        \\
            \toprule \multicolumn{1}{c}{}                 & \multicolumn{3}{c}{\textbf{Full images  (PSNR) $\uparrow$ }}       \\                                                                                                      
            \toprule
            HDD - Hard Patches                            & \cellcolor{green!30}56.31                                     & \cellcolor{green!30}55.30             & \cellcolor{green!30}54.64      \\
            HDD - All Patches                             & \cellcolor{yellow!35}55.85                                   & \cellcolor{yellow!35}54.80            & \cellcolor{yellow!35}53.97         \\
            HDD - Easy Patches                            & \cellcolor{blue!30}54.67                                       & \cellcolor{blue!30}53.78              & \cellcolor{blue!30}53.04              \\
        \end{tabular}
    }

    \label{tab:supp_esum_dataset}
\end{table*}

\begin{table*}[ht!]

	\caption[]{Results on hard patches and full images of different individual demosaicing methods on all three pattern types. We report results for BJDD~\cite{bjdd}, SAGAN~\cite{sagan}, and JDNDM~\cite{jdndm} on the pattern type they were designed for. We also report results for a version of JDNDM trained with shuffled Quad-Bayer and Nona-Bayer data. For remosaicing approaches (shuffle and DRUNet~\cite{drunet}), JDNDM is used as the Single-Bayer demosaicing network, and we report the size of the remosaicing model plus the size of JDNDM. In shuffle remosaicing, there is no model, so we only report the size of JDNDM. Results are reported at all 4 ISO levels in PSNR, SSIM, and Delta E. Green, yellow, and blue highlighting represents first, second, and third best, respectively.
	}
	\resizebox{1.0\textwidth}{!}{%
		\begin{tabular}{@{}lccccccccccccc}

			\toprule \multicolumn{1}{c}{}                         & \multicolumn{12}{c}{\textbf{Hard patches (PSNR)$\uparrow$}}                                                                                                                                                                                                                                                                                                                                                                                                                                               \\
			\toprule
			\multicolumn{1}{c}{}                                  & \multicolumn{1}{c|}{\multirow{2}{1.3cm}{\centering \textbf{ Size (MB) $\downarrow$ } }}{} & \multicolumn{3}{c|}{\textbf{ISO 400}} & \multicolumn{3}{c|}{\textbf{ISO 800 }} & \multicolumn{3}{c|}{\textbf{ISO 1600}} & \multicolumn{3}{c}{\textbf{ISO 3200}}                                                                                                                                                                                                                                               \\
			\multicolumn{1}{l}{\multirow{-1}{*}{\textbf{Method}}} & \multicolumn{1}{c|}{}                                                                     & \multicolumn{1}{c}{Single}            & Quad                                   & \multicolumn{1}{c|}{Nona}              & Single                                & Quad                        & \multicolumn{1}{c|}{Nona }  & Single                     & Quad                        & \multicolumn{1}{c|}{Nona}   & Single                     & Quad                        & Nona                        \\
			\midrule
			BJDD                                                  & 13.29                                                                                     & -                                     & \cellcolor{blue!30}50.86                & -                                      & -                                     & \cellcolor{blue!30} 50.05    & -                           & -                          & \cellcolor{blue!30}48.88     & -                           & -                          & 47.50                       & -                           \\

			SAGAN                                                 & 112.34                                                                                    & -                                     & -                                      & \cellcolor{blue!30} 49.55               & -                                     & -                           & \cellcolor{blue!30} 49.06    & -                          & -                           & \cellcolor{blue!30}48.05     & -                          & -                           & \cellcolor{blue!30}46.88     \\

			JDNDM                                                 & 12.21                                                                                     & \cellcolor{green!30}53.69             & -                                      & -                                      & \cellcolor{green!30}52.34             & -                           & -                           & \cellcolor{green!30}50.43  & -                           & -                           & \cellcolor{green!30}49.00  & -                           & -                           \\

			JDNDM  - Shuffle                                      & 12.21                                                                                     & -                                     & \cellcolor{green!30}52.15              & \cellcolor{green!30}51.32              & -                                     & \cellcolor{green!30}51.11   & \cellcolor{green!30}50.30   & -                          & \cellcolor{green!30}49.82   & \cellcolor{green!30}49.23   & -                          & \cellcolor{green!30}48.46   & \cellcolor{green!30}47.86   \\

			Remosaic Shuffle                                      & 12.21                                                                                     & -                                     & 40.42                                  & 36.48                                  & -                                     & 40.55                       & 36.60                       & -                          & 40.62                       & 36.73                       & -                          & 40.67                       & 36.85                       \\

			Remosaic DRUNet                                       & 148.81                                                                                    & -                                     & \cellcolor{yellow!35}51.78             & \cellcolor{yellow!35}51.00             & -                                     & \cellcolor{yellow!35}50.57  & \cellcolor{yellow!35}49.93  & -                          & \cellcolor{yellow!35}49.31  & \cellcolor{yellow!35}48.86  & -                          & \cellcolor{yellow!35}48.02  & \cellcolor{yellow!35}47.71  \\
			\toprule \multicolumn{1}{c}{}                         & \multicolumn{12}{c}{\textbf{Hard patches (SSIM)$\uparrow$}}                                                                                                                                                                                                                                                                                                                                                                                                                                               \\
			\midrule

			BJDD                                                  & 13.29                                                                                     & -                                     & \cellcolor{blue!30}0.9945               & -                                      & -                                     & \cellcolor{blue!30}0.9938    & -                           & -                          & \cellcolor{blue!30}0.9922    & -                           & -                          & \cellcolor{blue!30}0.9895    & -                           \\
			SAGAN                                                 & 112.34                                                                                    & -                                     & -                                      & \cellcolor{blue!30}0.9930               & -                                     & -                           & \cellcolor{blue!30}0.9926    & -                          & -                           & \cellcolor{blue!30}0.9908    & -                          & -                           & \cellcolor{blue!30}0.9884    \\
			JDNDM                                                 & 12.21                                                                                     & \cellcolor{green!30}0.9956            & -                                      & -                                      & \cellcolor{green!30}0.9957            & -                           & -                           & \cellcolor{green!30}0.9941 & -                           & -                           & \cellcolor{green!30}0.9908 & -                           & -                           \\
			JDNDM Shuffle                                         & 12.21                                                                                     & -                                     & \cellcolor{green!30}0.9958             & \cellcolor{green!30}0.9952             & -                                     & \cellcolor{green!30}0.9953  & \cellcolor{green!30}0.9944  & -                          & \cellcolor{green!30}0.9933  & \cellcolor{green!30}0.9926  & -                          & \cellcolor{green!30}0.9916  & \cellcolor{green!30}0.9902  \\
			Remosaic Shuffle                                      & 12.21                                                                                     & -                                     & 0.9656                                 & 0.9193                                 & -                                     & 0.9666                      & 0.9210                      & -                          & 0.9666                      & 0.9228                      & -                          & 0.9668                      & 0.9246                      \\
			Remosaic DRUNet                                       & 148.81                                                                                    & -                                     & \cellcolor{yellow!35}0.9953            & \cellcolor{yellow!35}0.9948            & -                                     & \cellcolor{yellow!35}0.9942 & \cellcolor{yellow!35}0.9933 & -                          & \cellcolor{yellow!35}0.9924 & \cellcolor{yellow!35}0.9919 & -                          & \cellcolor{yellow!35}0.9903 & \cellcolor{yellow!35}0.9898 \\
			\toprule \multicolumn{1}{c}{}                         & \multicolumn{12}{c}{\textbf{Hard patches (Delta E)$\downarrow$}}                                                                                                                                                                                                                                                                                                                                                                                                                                          \\
			\midrule
			BJDD                                                  & 13.29                                                                                     & -                                     & \cellcolor{blue!30}1.14                 & -                                      & -                                     & \cellcolor{blue!30}1.17      & -                           & -                          & \cellcolor{blue!30}1.29      & -                           & -                          & \cellcolor{blue!30}1.50      & -                           \\
			SAGAN                                                 & 112.34                                                                                    & -                                     & -                                      & \cellcolor{blue!30}1.30                 & -                                     & -                           & \cellcolor{blue!30}1.28      & -                          & -                           & \cellcolor{yellow!35}1.39   & -                          & -                           & \cellcolor{blue!30}1.57      \\
			JDNDM                                                 & 12.21                                                                                     & \cellcolor{green!30}0.96              & -                                      & -                                      & \cellcolor{green!30}0.93              & -                           & -                           & \cellcolor{green!30}1.15   & -                           & -                           & \cellcolor{green!30}1.43   & -                           & -                           \\
			JDNDM - Shuffle                                       & 12.21                                                                                     & -                                     & \cellcolor{yellow!35}0.98              & \cellcolor{green!30}1.02               & -                                     & \cellcolor{green!30}0.93    & \cellcolor{green!30}1.09    & -                          & \cellcolor{green!30}1.16    & \cellcolor{green!30}1.20    & -                          & \cellcolor{green!30}1.29    & \cellcolor{green!30}1.44    \\
			Remosaic Shuffle                                      & 12.21                                                                                     & -                                     & 1.86                                   & 2.87                                   & -                                     & 1.85                        & 2.84                        & -                          & 2.06                        & 2.98                        & -                          & 2.05                        & 2.94                        \\
			Remosaic DRUNet                                       & 148.81                                                                                    & -                                     & \cellcolor{green!30}0.94               & \cellcolor{green!30}1.02               & -                                     & \cellcolor{yellow!35}1.08   & \cellcolor{yellow!35}1.20   & -                          & \cellcolor{yellow!35}1.27   & \cellcolor{blue!30}1.41      & -                          & \cellcolor{yellow!35}1.42   & \cellcolor{yellow!35}1.50   \\

			\midrule
			\\
			\toprule \multicolumn{1}{c}{}                         & \multicolumn{12}{c}{\textbf{Full images (PSNR)$\uparrow$}}                                                                                                                                                                                                                                                                                                                                                                                                                                                \\
			\midrule
			BJDD                                                  & 13.29                                                                                     & -                                     & \cellcolor{blue!30}53.47                & -                                      & -                                     & \cellcolor{blue!30} 52.72    & -                           & -                          & \cellcolor{blue!30}51.63     & -                           & -                          & \cellcolor{blue!30} 50.16    & -                           \\

			SAGAN                                                 & 112.34                                                                                    & -                                     & -                                      & \cellcolor{blue!30} 51.37               & -                                     & -                           & \cellcolor{blue!30}51.62     & -                          & -                           & \cellcolor{blue!30}50.72     & -                          & -                           & \cellcolor{blue!30}49.42     \\

			JDNDM                                                 & 12.21                                                                                     & \cellcolor{green!30}56.36             & -                                      & -                                      & \cellcolor{green!30}55.01             & -                           & -                           & \cellcolor{green!30}53.04  & -                           & -                           & \cellcolor{green!30}51.74  & -                           & -                           \\
			JDNDM  - Shuffle                                      & 12.21                                                                                     & -                                     & \cellcolor{green!30}54.70              & \cellcolor{green!30}54.03              & -                                     & \cellcolor{green!30}53.75   & \cellcolor{green!30}53.07   & -                          & \cellcolor{green!30}52.52   & \cellcolor{green!30}52.06   & -                          & \cellcolor{green!30}51.19   & \cellcolor{green!30}50.64   \\

			Remosaic Shuffle                                      & 12.21                                                                                     & -                                     & 43.53                                  & 39.37                                  & -                                     & 43.65                       & 39.48                       & -                          & 43.75                       & 39.64                       & -                          & 43.78                       & 39.74                       \\

			Remosaic DRUNet                                       & 148.81                                                                                    & -                                     & \cellcolor{yellow!35}54.59             & \cellcolor{yellow!35}53.76             & -                                     & \cellcolor{yellow!35}53.16  & \cellcolor{yellow!35}52.57  & -                          & \cellcolor{yellow!35}51.85  & \cellcolor{yellow!35}51.39  & -                          & \cellcolor{yellow!35}50.72  & \cellcolor{yellow!35}50.06  \\
			\toprule \multicolumn{1}{c}{}                         & \multicolumn{12}{c}{\textbf{Full images (SSIM)$\uparrow$}}                                                                                                                                                                                                                                                                                                                                                                                                                                                \\
			\midrule
			BJDD                                                  & 13.29                                                                                     & -                                     & \cellcolor{blue!30}0.9964               & -                                      & -                                     & \cellcolor{blue!30}0.9964    & -                           & -                          & \cellcolor{yellow!35}0.9955 & -                           & -                          & \cellcolor{blue!30}0.9932    & -                           \\
			SAGAN                                                 & 112.34                                                                                    & -                                     & -                                      & \cellcolor{blue!30}0.9942               & -                                     & -                           & \cellcolor{blue!30}0.9954    & -                          & -                           & \cellcolor{blue!30}0.9947    & -                          & -                           & \cellcolor{blue!30}0.9928    \\
			JDNDM                                                 & 12.21                                                                                     & \cellcolor{green!30}0.9980            & -                                      & -                                      & \cellcolor{green!30}0.9979            & -                           & -                           & \cellcolor{green!30}0.9968 & -                           & -                           & \cellcolor{green!30}0.9952 & -                           & -                           \\
			JDNDM - Shuffle                                       & 12.21                                                                                     & -                                     & \cellcolor{yellow!35}0.9975            & \cellcolor{green!30}0.9973             & -                                     & \cellcolor{green!30}0.9977  & \cellcolor{green!30}0.9970  & -                          & \cellcolor{green!30}0.9963  & \cellcolor{green!30}0.9961  & -                          & \cellcolor{green!30}0.9958  & \cellcolor{green!30}0.9942  \\
			Remosaic Shuffle                                      & 12.21                                                                                     & -                                     & 0.9859                                 & 0.9655                                 & -                                     & 0.9863                      & 0.9664                      & -                          & 0.9855                      & 0.9667                      & -                          & 0.9861                      & 0.9680                      \\
			Remosaic DRUNet                                       & 148.81                                                                                    & -                                     & \cellcolor{green!30}0.9977             & \cellcolor{green!30}0.9973             & -                                     & \cellcolor{yellow!35}0.9966 & \cellcolor{yellow!35}0.9961 & -                          & \cellcolor{blue!30}0.9952    & \cellcolor{yellow!35}0.9950 & -                          & \cellcolor{yellow!35}0.9947 & \cellcolor{yellow!35}0.9939 \\

			\toprule \multicolumn{1}{c}{}                         & \multicolumn{12}{c}{\textbf{Full images (Delta E)$\downarrow$}}                                                                                                                                                                                                                                                                                                                                                                                                                                           \\
			\midrule
			BJDD                                                  & 13.29                                                                                     & -                                     & \cellcolor{blue!30}1.20                 & -                                      & -                                     & \cellcolor{blue!30}1.28      & -                           & -                          & \cellcolor{yellow!35}1.33   & -                           & -                          & \cellcolor{blue!30}1.59      & -                           \\
			SAGAN                                                 & 112.34                                                                                    & -                                     & -                                      & \cellcolor{blue!30}1.66                 & -                                     & -                           & \cellcolor{blue!30}1.46      & -                          & -                           & \cellcolor{yellow!35}1.46   & -                          & -                           & \cellcolor{blue!30}1.87      \\
			JDNDM                                                 & 12.21                                                                                     & \cellcolor{green!30}0.89              & -                                      & -                                      & \cellcolor{green!30}0.86              & -                           & -                           & \cellcolor{green!30}1.14   & -                           & -                           & \cellcolor{green!30}1.38   & -                           & -                           \\
			JDNDM - Shuffle                                       & 12.21                                                                                     & -                                     & \cellcolor{yellow!35}1.05              & \cellcolor{yellow!35}1.06              & -                                     & \cellcolor{green!30}0.85    & \cellcolor{green!30}1.07    & -                          & \cellcolor{green!30}1.16    & \cellcolor{green!30}1.15    & -                          & \cellcolor{green!30}1.21    & \cellcolor{green!30}1.48    \\
			Remosaic Shuffle                                      & 12.21                                                                                     & -                                     & 1.46                                   & 2.16                                   & -                                     & 1.50                        & 2.15                        & -                          & 1.88                        & 2.47                        & -                          & 1.74                        & 2.32                        \\
			Remosaic DRUNet                                       & 148.81                                                                                    & -                                     & \cellcolor{green!30}0.90               & \cellcolor{green!30}1.02               & -                                     & \cellcolor{yellow!35}1.17   & \cellcolor{yellow!35}1.29   & -                          & \cellcolor{blue!30}1.40      & \cellcolor{blue!30}1.51      & -                          & \cellcolor{yellow!35}1.41   & \cellcolor{yellow!35}1.52   \\
			\bottomrule
		\end{tabular}
	}

	\label{tab:full_individual_models}
\end{table*}

\begin{table*}[h!] \
	\small
	\centering

	\caption{Results on hard patches and full images of our three types of unified models and KLAP~\cite{KLAP}. For SRUM and LSUM, we try models with both shuffle heads and packing heads. Results are reported at all 4 ISO levels in PSNR, SSIM, and Delta E. Green, yellow, and blue highlighting represents first, second, and third best, respectively. }

	\resizebox{1.0\textwidth}{!}{%
		\begin{tabular}{@{}lccccccccccccc}
			\toprule \multicolumn{1}{c}{}                         & \multicolumn{12}{c}{\textbf{Hard patches (PSNR)$\uparrow$}}                                                                                                                                                                                                                                                                                                                                                                                                                                               \\

			\toprule
			\multicolumn{1}{c}{}                                  & \multicolumn{1}{c|}{\multirow{2}{1.3cm}{\centering \textbf{ Size (MB) $\downarrow$ } }}{} & \multicolumn{3}{c|}{\textbf{ISO 400}} & \multicolumn{3}{c|}{\textbf{ISO 800 }} & \multicolumn{3}{c|}{\textbf{ISO 1600}} & \multicolumn{3}{c}{\textbf{ISO 3200}}                                                                                                                                                                                                                                                    \\
			\multicolumn{1}{l}{\multirow{-1}{*}{\textbf{Method}}} & \multicolumn{1}{c|}{}                                                                     & \multicolumn{1}{c}{Single}            & Quad                                   & \multicolumn{1}{c|}{Nona}              & Single                                & Quad                        & \multicolumn{1}{c|}{Nona }  & Single                       & Quad                         & \multicolumn{1}{c|}{Nona}    & Single                      & Quad                        & Nona                        \\
			\toprule
			KLAP                                                  & 25.62                                                                                     & 53.27                                 & 52.01                                  & 50.44                                  & 51.91                                 & 50.99                       & 49.79                       & 50.28                        & 49.40                        & 48.59                        & \cellcolor{blue!30}48.95     & 48.10                       & 47.20                       \\

			SRUM Shuffle                                          & 14.01                                                                                    & 53.33                                 & 51.83                                  & 50.92                                  & \cellcolor{blue!30}52.10               & 50.89                       & 50.24                       & 50.35                        & 49.46                        & 48.99                        & 48.91                       & 48.19                       & 47.76                       \\
			SRUM Packing                                          & 16.53                                                                                     & 53.12                                 & 51.52                                  & 50.58                                  & 51.70                                 & 50.39                       & 49.70                       & \cellcolor{blue!30}50.36      & 49.16                        & 48.82                        & 48.87                       & 48.25                       & 47.71                       \\
			LSUM Shuffle                                          & 13.41                                                                                     & \cellcolor{blue!30}53.47               & \cellcolor{yellow!35}52.38             & \cellcolor{yellow!35}51.58             & \cellcolor{green!25}52.17             & \cellcolor{blue!30}51.16     & \cellcolor{yellow!35}50.63  & 50.28                        & \cellcolor{yellow!35}49.84   & \cellcolor{yellow!35}49.13   & \cellcolor{green!25}49.03   & \cellcolor{green!25}48.58   & \cellcolor{yellow!35}48.09  \\
			LSUM Packing                                          & 15.92                                                                                     & \cellcolor{yellow!35}53.55            & \cellcolor{blue!30}52.27                & \cellcolor{blue!30}51.37                & 52.08                                 & \cellcolor{yellow!35}51.22  & \cellcolor{blue!30}50.42     & \cellcolor{yellow!35}50.50   & \cellcolor{blue!30}49.71      & \cellcolor{blue!30}49.03      & 48.93                       & \cellcolor{blue!30}48.38     & \cellcolor{blue!30}47.77     \\
			ESUM                                                  & 12.21                                                                                     & \cellcolor{green!25}53.75             & \cellcolor{green!25}52.68              & \cellcolor{green!25}51.96              & 52.17\cellcolor{green!25}             & 51.36\cellcolor{green!25}   & 50.76\cellcolor{green!25}   & \cellcolor{green!25}50.64    & \cellcolor{green!25}50.01    & \cellcolor{green!25}49.46    & \cellcolor{yellow!35}48.98  & \cellcolor{yellow!35}48.57  & \cellcolor{green!25}48.11   \\
			\toprule \multicolumn{1}{c}{}                         & \multicolumn{12}{c}{\textbf{Hard patches (SSIM)$\uparrow$}}                                                                                                                                                                                                                                                                                                                                                                                                                                                    \\
			\midrule
			KLAP                                                  & 25.62                                                                                     & \cellcolor{blue!30}0.9969              & \cellcolor{blue!30}0.9958               & 0.9943                                 & \cellcolor{blue!30}0.9959              & \cellcolor{blue!30}0.9948    & 0.9935                      & \cellcolor{yellow!35}0.9942 & 0.9929                       & 0.9917                       & \cellcolor{blue!30}0.9922    & 0.9907                      & 0.9889                      \\
			SRUM Shuffle                                          & 14.01                                                                                    & \cellcolor{blue!30}0.9969              & 0.9957                                 & 0.9949                                 & \cellcolor{blue!30}0.9959              & 0.9946                      & 0.9940                      & 0.9941                       & 0.9928                       & \cellcolor{yellow!35}0.9924 & \cellcolor{blue!30}0.9922    & 0.9910                      & \cellcolor{blue!30}0.9901    \\
			SRUM Packing                                          & 16.53                                                                                     & 0.9968                                & 0.9952                                 & 0.9941                                 & 0.9955                                & 0.9940                      & 0.9933                      & \cellcolor{yellow!35}0.9942 & 0.9921                       & 0.9918                       & 0.9920                      & 0.9909                      & 0.9899                      \\
			LSUM Shuffle                                          & 13.41                                                                                     & \cellcolor{blue!30}0.9969              & \cellcolor{yellow!35}0.9959            & \cellcolor{yellow!35}0.9953            & \cellcolor{yellow!35}0.9960           & \cellcolor{yellow!35}0.9950 & \cellcolor{yellow!35}0.9945 & 0.9941                       & \cellcolor{yellow!35}0.9934 & \cellcolor{yellow!35}0.9924 & \cellcolor{yellow!35}0.9923 & \cellcolor{yellow!35}0.9915 & \cellcolor{yellow!35}0.9906 \\
			LSUM Packing                                          & 15.92                                                                                     & \cellcolor{yellow!35}0.9970           & \cellcolor{yellow!35}0.9959            & \cellcolor{blue!30}0.9952               & \cellcolor{blue!30}0.9959              & \cellcolor{yellow!35}0.9950 & \cellcolor{blue!30}0.9942    & \cellcolor{yellow!35}0.9942 & \cellcolor{blue!30}0.9932     & 0.9921                       & \cellcolor{blue!30}0.9922    & \cellcolor{blue!30}0.9912    & 0.9899                      \\
			ESUM                                                  & 12.21                                                                                     & \cellcolor{green!25}0.9971            & \cellcolor{green!25}0.9962             & \cellcolor{green!25}0.9957             & \cellcolor{green!25}0.9961            & \cellcolor{green!25}0.9954  & \cellcolor{green!25}0.9948 & \cellcolor{green!25}0.9945   & \cellcolor{green!25}0.9936   & \cellcolor{green!25}0.9929   & \cellcolor{green!25}0.9924  & \cellcolor{green!25}0.9917  & \cellcolor{green!25}0.9909  \\
			\toprule \multicolumn{1}{c}{}                         & \multicolumn{12}{c}{\textbf{Hard patches (Delta E)$\downarrow$}}                                                                                                                                                                                                                                                                                                                                                                                                                                               \\
			\midrule
			KLAP                                                  & 25.62                                                                                     & \cellcolor{blue!30}0.85                & \cellcolor{yellow!35}0.93              & 1.16                                   & \cellcolor{yellow!35}0.96             & \cellcolor{yellow!35}1.01   & \cellcolor{blue!30}1.15      & \cellcolor{yellow!35}1.13    & 1.25                         & 1.29                         & \cellcolor{green!25}1.27    & 1.47                        & 1.53                        \\
			SRUM Shuffle                                          & 14.01                                                                                    & 0.88                                  & 1.01                                   & 1.12                                   & \cellcolor{yellow!35}0.96             & 1.10                        & 1.18                        & 1.16                         & 1.38                         & \cellcolor{blue!30}1.28       & \cellcolor{blue!30}1.29      & \cellcolor{blue!30}1.38      & \cellcolor{blue!30}1.46      \\
			SRUM Packing                                          & 16.53                                                                                     & 0.86                                  & 1.04                                   & 1.21                                   & 1.07                                  & 1.15                        & 1.23                        & 1.15                         & 1.40                         & 1.43                         & 1.37                        & 1.46                        & 1.56                        \\
			LSUM Shuffle                                          & 13.41                                                                                     & 0.89                                  & 0.96                                   & \cellcolor{yellow!35}1.03              & \cellcolor{yellow!35}0.96             & \cellcolor{blue!30}1.05      & \cellcolor{yellow!35}1.07   & 1.33                         & \cellcolor{yellow!35}1.20    & \cellcolor{yellow!35}1.25    & \cellcolor{yellow!35}1.28   & \cellcolor{yellow!35}1.32   & \cellcolor{yellow!35}1.38   \\
			LSUM Packing                                          & 15.92                                                                                     & \cellcolor{yellow!35}0.84             & \cellcolor{yellow!35}0.93              & \cellcolor{blue!30}1.06                 & 1.04                                  & 1.08                        & 1.20                        & \cellcolor{yellow!35}1.13    & \cellcolor{blue!30}1.24       & 1.36                         & 1.33                        & \cellcolor{blue!30}1.38      & 1.48                        \\
			ESUM                                                  & 12.21                                                                                     & \cellcolor{green!25}0.80              & \cellcolor{green!25}0.88               & \cellcolor{green!25}0.94               & \cellcolor{green!25}0.95              & \cellcolor{green!25}0.95    & \cellcolor{green!25}1.00    & \cellcolor{green!25}1.08     & \cellcolor{green!25}1.14     & \cellcolor{green!25}1.21     & \cellcolor{blue!30}1.29      & \cellcolor{green!25}1.30    & \cellcolor{green!25}1.35    \\

			\midrule
			\\
			\toprule \multicolumn{1}{c}{}                         & \multicolumn{12}{c}{\textbf{Full images (PSNR)$\uparrow$}}                                                                                                                                                                                                                                                                                                                                                                                                                                                     \\
			\midrule
			KLAP                                                  & 25.62                                                                                     & 55.84                                 & 54.68                                  & 52.70                                  & 54.41                                 & 53.65                       & 52.42                       & 52.96                        & 52.05                        & 51.11                        & 51.49                       & 50.68                       & 49.61                       \\

			SRUM Shuffle                                          & 14.01                                                                                    & 55.93                                 & 54.45                                  & 53.62                                  & \cellcolor{yellow!35}54.76            & 53.35                       & 52.78                       & \cellcolor{blue!30}53.04      & 52.01                        & \cellcolor{blue!30}51.80      & 51.63                       & 50.84                       & 50.48                       \\
			SRUM Packing                                          & 16.53                                                                                     & 55.74                                 & 54.14                                  & 53.23                                  & 54.29                                 & 53.02                       & 52.44                       & 53.03                        & 51.73                        & 51.54                        & 51.59                       & 51.03                       & \cellcolor{blue!30}50.56     \\
			LSUM Shuffle                                          & 13.41                                                                                     & \cellcolor{blue!30}56.09               & \cellcolor{blue!30}54.79                & \cellcolor{blue!30}53.99                & \cellcolor{green!25}54.82             & \cellcolor{blue!30}53.72     & \cellcolor{yellow!35}53.23  & 52.95                        & \cellcolor{yellow!35}52.47   & \cellcolor{yellow!35}51.82   & \cellcolor{green!25}51.78   & \cellcolor{yellow!35}51.26  & \cellcolor{yellow!35}50.86  \\
			LSUM Packing                                          & 15.92                                                                                     & \cellcolor{yellow!35}56.19            & \cellcolor{yellow!35}55.01             & \cellcolor{yellow!35}54.13             & 54.65                                 & \cellcolor{yellow!35}53.93  & \cellcolor{blue!30}53.10     & \cellcolor{yellow!35}53.17   & \cellcolor{blue!30}52.42      & 51.69                        & \cellcolor{blue!30}51.66     & \cellcolor{blue!30}51.14     & 50.50                       \\
			ESUM                                                  & 12.21                                                                                     & \cellcolor{green!25}56.31             & \cellcolor{green!25}55.30              & \cellcolor{green!25}54.64              & 54.71\cellcolor{blue!30}               & 53.97\cellcolor{green!25}   & 53.43\cellcolor{green!25}   & \cellcolor{green!25}53.35    & \cellcolor{green!25}52.76    & \cellcolor{green!25}52.25    & \cellcolor{yellow!35}51.69  & \cellcolor{green!25}51.34   & \cellcolor{green!25}50.93   \\
			\toprule \multicolumn{1}{c}{}                         & \multicolumn{12}{c}{\textbf{Full images (SSIM)$\uparrow$}}                                                                                                                                                                                                                                                                                                                                                                                                                                                     \\
			\midrule
			KLAP                                                  & 25.62                                                                                     & \cellcolor{blue!30}0.9983              & \cellcolor{yellow!35}0.9979            & 0.9966                                 & 0.9976                                & \cellcolor{yellow!35}0.9973 & 0.9965                      & \cellcolor{yellow!35}0.9968  & \cellcolor{yellow!35}0.9961  & 0.9952                       & \cellcolor{blue!30}0.9957    & 0.9943                      & 0.9934                      \\
			SRUM Shuffle                                          & 14.01                                                                                    & \cellcolor{blue!30}0.9983              & 0.9977                                 & \cellcolor{blue!30}0.9971               & \cellcolor{green!25}0.9978            & 0.9963                      & 0.9963                      & \cellcolor{blue!30}0.9966     & 0.9952                       & \cellcolor{yellow!35}0.9957  & \cellcolor{green!25}0.9958  & 0.9946                      & 0.9941                      \\
			SRUM Packing                                          & 16.53                                                                                     & 0.9982                                & 0.9972                                 & 0.9965                                 & 0.9970                                & 0.9961                      & 0.9960                      & \cellcolor{blue!30}0.9966     & 0.9946                       & 0.9952                       & 0.9954                      & 0.9949                      & \cellcolor{blue!30}0.9945    \\
			LSUM Shuffle                                          & 13.41                                                                                     & 0.9981                                & 0.9974                                 & 0.9965                                 & \cellcolor{green!25}0.9978            & 0.9970                      & \cellcolor{yellow!35}0.9968 & 0.9964                       & \cellcolor{blue!30}0.9960     & \cellcolor{blue!30}0.9953     & \cellcolor{green!25}0.9958  & \cellcolor{yellow!35}0.9953 & \cellcolor{yellow!35}0.9949 \\
			LSUM Packing                                          & 15.92                                                                                     & \cellcolor{green!25}0.9984            & \cellcolor{yellow!35}0.9979            & \cellcolor{yellow!35}0.9974            & 0.9974                                & \cellcolor{blue!30}0.9972    & \cellcolor{blue!30}0.9966    & \cellcolor{blue!30}0.9966     & \cellcolor{blue!30}0.9960     & 0.9949                       & 0.9956                      & \cellcolor{blue!30}0.9951    & 0.9938                      \\
			ESUM                                                  & 12.21                                                                                     & \cellcolor{green!25}0.9984            & \cellcolor{green!25}0.9980             & \cellcolor{green!25}0.9978             & \cellcolor{green!25}0.9978            & \cellcolor{green!25}0.9976  & \cellcolor{green!25}0.9974  & \cellcolor{green!25}0.9970   & \cellcolor{green!25}0.9966   & \cellcolor{green!25}0.9962   & 0.9956                      & \cellcolor{green!25}0.9955  & \cellcolor{green!25}0.9953  \\
			\toprule \multicolumn{1}{c}{}                         & \multicolumn{12}{c}{\textbf{Full images (Delta E)$\downarrow$}}                                                                                                                                                                                                                                                                                                                                                                                                                                                \\
			\midrule
			KLAP                                                  & 25.62                                                                                     & \cellcolor{yellow!35}0.84             & \cellcolor{yellow!35}0.89              & 1.19                                   & \cellcolor{blue!30}1.02                & \cellcolor{yellow!35}1.00   & \cellcolor{blue!30}1.16      & \cellcolor{blue!30}1.16       & \cellcolor{blue!30}1.27       & \cellcolor{blue!30}1.34       & \cellcolor{yellow!35}1.28   & 1.63                        & 1.63                        \\
			SRUM Shuffle                                          & 14.01                                                                                    & 0.91                                  & 0.99                                   & \cellcolor{blue!30}1.13                 & 0.98                                  & 1.22                        & 1.29                        & 1.21                         & 1.60                         & 1.30                         & \cellcolor{green!25}1.27    & 1.44                        & \cellcolor{blue!30}1.51      \\
			SRUM Packing                                          & 16.53                                                                                     & 0.87                                  & 1.06                                   & 1.27                                   & 1.18                                  & 1.15                        & 1.21                        & \cellcolor{yellow!35}1.15    & 1.59                         & 1.54                         & 1.41                        & 1.49                        & 1.53                        \\
			LSUM Shuffle                                          & 13.41                                                                                     & 0.92                                  & 1.07                                   & 1.27                                   & \cellcolor{green!25}1.00              & \cellcolor{blue!30}1.10      & \cellcolor{yellow!35}1.14   & 1.46                         & \cellcolor{yellow!35}1.25    & \cellcolor{yellow!35}1.23    & \cellcolor{green!25}1.27    & \cellcolor{yellow!35}1.30   & \cellcolor{yellow!35}1.38   \\
			LSUM Packing                                          & 15.92                                                                                     & \cellcolor{blue!30}0.86                & \cellcolor{blue!30}0.94                 & \cellcolor{yellow!35}1.08              & 1.16                                  & 1.12                        & 1.29                        & 1.18                         & 1.28                         & 1.45                         & 1.38                        & \cellcolor{blue!30}1.41      & 1.55                        \\
			ESUM                                                  & 12.21                                                                                     & \cellcolor{green!25}0.82              & \cellcolor{green!25}0.88               & \cellcolor{green!25}0.93               & \cellcolor{yellow!35}1.01             & \cellcolor{green!25}0.92    & \cellcolor{green!25}0.97    & \cellcolor{green!25}1.09     & \cellcolor{green!25}1.15     & \cellcolor{green!25}1.21     & 1.34                        & \cellcolor{green!25}1.29    & \cellcolor{green!25}1.32    \\
			\bottomrule
		\end{tabular}
	}

	\label{tab:full_unified_models}
\end{table*}

\newpage
\clearpage
\clearpage


\end{document}